\documentclass[reprint,amsmath,amssymb,aps,nofootinbib]{revtex4-2}
\usepackage{graphicx}
\usepackage{dcolumn}
\usepackage{bm}
\usepackage{physics}
\usepackage{hyperref}
\hypersetup{
     colorlinks=true,
     linkcolor=blue,
     filecolor=blue,
     citecolor = blue,      
     urlcolor=blue,
     }
\usepackage[super]{nth}
\usepackage{physics}

\usepackage[shortlabels]{enumitem}
\usepackage{bbm}
\usepackage{orcidlink}
\usepackage{tensor}

\newcommand{\gi}[1]{\ensuremath{g^{(#1)}_{\mu\nu}}}

\newcommand{\Wi}[1]{\ensuremath{W^{(#1)\mu}_{\;\;\;\;\;\;\nu}}}
\newcommand{\Gi}[1]{\ensuremath{G^{(#1)\mu}_{\;\;\;\;\;\;\nu}}}

\newcommand{\Tmunui}[1]{\ensuremath{T^{(#1)\mu}_{\;\;\;\;\;\;\nu}}}
\newcommand{\vbein}[3]{\ensuremath{e_{#2}^{(#1)#3}}}
\newcommand{\ivbein}[3]{\ensuremath{e_{\;\;\;\;\;#3}^{(#1)#2}}}

\newcommand{\Sij}[2]{\ensuremath{S_{#1\rightarrow#2}}}

\newcommand{\dg}[3]{\ensuremath{\tensor{\delta g}{^{(#1)#2}_#3}}}
\newcommand{\om}[3]{\ensuremath{\tensor{\omega}{^{(#1)#2}_#3}}}

\newcommand{\Tcoeffs}{\ensuremath{T_{j_1\hdots j_D}}}

\newcommand{\hodge}{\ensuremath{\star}}

\newcommand{\gdelta}[1]{\ensuremath{\delta^{\mu_1\hdots\mu_#1}_{\nu_1\hdots\nu_#1}}}

\begin{document}

\title{New formalism for perturbations of massive gravity theories around arbitrary background spacetimes}
\author{Kieran Wood \orcidlink{0000-0002-4680-5563}}
\email{kieran.wood@nottingham.ac.uk}
\affiliation{School of Physics and Astronomy, University Of Nottingham, Nottingham NG7 2RD, UK}
\affiliation{Nottingham Centre Of Gravity, University Of Nottingham, Nottingham NG7 2RD, UK}

\begin{abstract}
    We develop a new technique for studying the perturbations of dRGT-type massive gravity theories around arbitrary background spacetimes. Built initially from the vielbein formulation of the theory, but switching back to the metric formulation afterwards, our approach bypasses many of the complications that arise in previous metric formulation approaches to linearising massive gravity around generic backgrounds, naturally elucidates the ghost-free structure of the interactions, and readily generalises to higher orders in perturbation theory, as well as to multiple interacting metric tensor fields. To demonstrate the power of our technique, we apply our formalism to a number of commonly occurring example backgrounds -- proportional, cosmological, and black hole -- recovering and extending many known results from the literature at linear order. Lastly, we provide, for the first time, the cubic order multi-gravity potential around a generic background spacetime.
\end{abstract}

\maketitle

\section{Introduction}\label{sec:intro}

Recent years have seen something of a resurgence in the study of massive spin-2 fields and their interactions, owing to many interesting theoretical developments, as well as potential applications to a number of outstanding fundamental physics problems, particularly in cosmology (see \cite{Hinterbichler_review,dR_review,bigravity_review,KWThesis} for reviews). The subject has a rich history dating back to the time of Fierz and Pauli (FP), who in 1939 first wrote down the only consistent linear theory of such a field propagating on Minkowski spacetime \cite{FierzPauli}, later extended to also apply for generic Einstein spacetimes \cite{OG_Higuchi,OG_Higuchi_2} (i.e. those of constant curvature). The FP theory is essentially linearised general relativity (GR) -- the unique theory of a \emph{massless} spin-2 field -- supplemented by a mass term containing all possible quadratic contractions of the metric perturbation $h_{\mu\nu}$ i.e.
\begin{equation}\label{FP}
    \mathcal{L}_{\text{FP}} \supset -\frac{m^2}{8}\left(h_{\mu\nu}h^{\mu\nu} + A h^2\right) \; ,
\end{equation}
with $h=\bar{g}^{\mu\nu}h_{\mu\nu}$ the trace with respect to the background metric $\bar{g}_{\mu\nu}$. Such a mass term, if $A$ is chosen arbitrarily, generically excites a ghostly scalar mode contained within $h_{\mu\nu}$ -- the so-called Boulware-Deser (BD) ghost -- whose presence would render the vacuum of the theory unstable upon coupling $h_{\mu\nu}$ to matter. The magic of the FP choice of $A=-1$ is that it ensures $h_{00}$ appears in the action as a Lagrange multiplier, enforcing a primary constraint. A further secondary constraint arises from the assertion that the primary constraint should be preserved in time, and together the two constraints act to exorcise the ghost mode and its conjugate momentum from the spectrum of the theory, leaving 5 physical, healthy degrees of freedom in 4 dimensions.

Extending the linear FP theory to a fully nonlinear theory of \emph{massive gravity} (i.e. GR plus a nonlinear mass term) was long thought to be an impossible task, as most candidate nonlinear mass terms are doomed to resurrect the BD ghost \cite{BD_Ghosts,BD_ghost_explicit}. It was only relatively recently, in 2010, that a satisfactory mass term (in fact, the \emph{unique} such term) circumventing Boulware and Deser's apparent no-go result was found, and subsequently proved to be free from the vexatious ghost at the full nonlinear level \cite{dRGT_1,dRGT_2,first_action_metric_form,ghost_freedom_flat_ref,ghost_freedom_general_ref,ghost_freedom_stuckelberg,hamiltonian_analysis,Kluson,Kluson_note,Covariant_approach_no_ghosts}. Built upon groundwork laid earlier in \cite{EFTforMGs,Creminelli}, the associated theory of gravity now goes by the name \emph{dRGT massive gravity}, after its three original architects: de Rham, Gabadadze and Tolley, although the two principal formulations with which we typically express it today (metric and vielbein) are actually due to Hassan, Rosen and Hinterbichler \cite{first_action_metric_form,interacting_spin2}.

The nonlinear mass term in dRGT theory is constructed from an interaction between \emph{two} independent metrics: the physical metric of spacetime, and some fiducial, non-dynamical reference metric that one inserts by hand (typically taken to be Minkowski, but one is free to be more general if they so wish). By providing a kinetic term for the reference metric, thereby promoting it to a second dynamical field, one obtains the theory of \emph{bigravity} \cite{HR1}, which, due to the special structure of the dRGT interactions, is also ghost free \cite{HR2,official_secondary_constraint_bigravity}. The generalisation to multiple interacting metrics came soon after in \cite{interacting_spin2} (see also \cite{Multi_via_massive}), although the general \emph{multi-gravity} theory is only devoid of the BD ghost up to certain conditions \cite{3D_vbein_nocycles,ghost_freedom_multigravity,cycles,beyond_pairwise_couplings}, upon which we shall elaborate in section \ref{sec:multigrav} when we cover the mathematics of the theory in detail. 

In pure dRGT massive gravity, the fixed nature of the reference metric means that the would-be diffeomorphism invariance of GR is completely broken by the mass term and hence the one propagating spin-2 field in the theory has a mass. However, in bi- and multi-gravity, the interaction term remains invariant under the diagonal subgroup of diffeomorphisms that transforms every metric in the same way, hence these theories contain a single massless spin-2 field (associated to this diagonal subgroup) as well as a collection of massive spin-2 fields (associated to the broken diffeomorphisms). This is consistent with a powerful no-go theorem by Boulanger, Damour, Gualtieri and Henneaux stating, up to some mild assumptions\footnote{Namely: locality, compatibility with Poincaré invariance, spacetime dimension $D>2$ and a Lagrangian that is at most second-order in derivatives.}, that theories containing multiple \emph{massless} spin-2 fields interacting nonlinearly are inconsistent \cite{no_interacting_massless_gravitons}: in any theory of interacting spin-2 fields, all but one of them must be massive. 

As expected, dRGT massive gravity correctly linearises to the FP theory around an Einstein background. Multi-gravity theories linearise to a sort of `multi-FP' theory containing a mass \emph{matrix} coupling different metric perturbations at quadratic order (in this sense the metrics themselves are akin to flavour eigenstates) i.e.
\begin{equation}\label{MultiFP}
    \mathcal{L}_{\text{mFP}} \supset -\frac{\mathcal{M}_{ij}^2}{8}\left[h^{(i)}_{\mu\nu}h^{(j)\mu\nu}-h^{(i)}h^{(j)}\right] \; ,
\end{equation}
around backgrounds where all metrics are proportional to the same Einstein space, as we will see explicitly in section \ref{sec:examples}. The matrix $\mathcal{M}_{ij}^2$ always contains precisely one zero eigenvalue, in accordance with the no-go theorem mentioned earlier \cite{consistent_spin2,prop_bg_multigrav,BH2}. 

Around more generic backgrounds, the structure of the perturbations is much more abstruse. This is a problem, because many spacetimes of physical relevance are \emph{not} Einstein spaces, and metric perturbations around them can give rise to important physical effects e.g. around the FLRW metric of cosmology, metric perturbations seed the growth of structure in the universe. Thus, if one wishes to study such effects in the context of these massive gravity theories, it remains an important task to figure out how to perturb them around arbitrary backgrounds in a systematic manner. Thankfully, we already have a pretty good handle on how to do this in the metric formulation of multi-gravity up to linear order, owing initially to a series of papers by Bernard, Deffayet, Schmidt-May and von Strauss \cite{MG_arbitrary_BG,Syzygies,linear_spin2_gen_BG} from 2015, followed by the papers \cite{Volkov_arbitraryBG,Volkov_arbitraryBG_detailed} by Mazuet and Volkov from a couple of years later (see also \cite{Durrer_perts_1,general_mass}). The procedure of \cite{MG_arbitrary_BG,Syzygies,linear_spin2_gen_BG}, in its original form, is formulaic and readily applicable in principle, but the calculations involved are complex, containing many steps that can quickly become tedious around more complicated backgrounds, and that do not easily generalise to higher orders in perturbation theory. The reason for this, as we will see in section \ref{sec:multigrav}, is that the interaction term in the metric formulation contains a matrix square root, which can be very awkward to handle, especially at the level of the perturbations: even at linear order, to determine the structure of the perturbed mass term one is forced to solve a complicated matrix equation, whose solution can be highly non-trivial \cite{Sylvester_matrix_eq}. The authors did recognise this issue, and provided a means to sidestep it in \cite{linear_spin2_gen_BG}, at least for dRGT massive gravity and bigravity, by using some clever field redefinitions. Essentially, by absorbing the \emph{background} square root matrix into the definition of the metric perturbations, it becomes possible to write down the linearised field equations in terms of these new variables without having to solve any complicated matrix equations. They used this approach to demonstrate the existence of the ghost-killing constraints (at linear level) around generic backgrounds in bigravity in a covariant manner. However, the necessary field redefinitions inevitably muddy the kinetic structure of the perturbations (this is really only an aesthetic issue), and while their procedure works a treat for theories with exactly 2 metrics, where there is only a single interaction, it will no longer work for theories containing more than 2 interacting metrics (this is a more serious issue)  -- we will explain why in appendix \ref{app:old perts}. 

The work of \cite{Volkov_arbitraryBG,Volkov_arbitraryBG_detailed} improved on this situation by utilising the equivalence between the vielbein and metric formulations of dRGT massive gravity to sidestep the necessity of dealing with the troublesome matrix square roots, instead working directly with the perturbations of the vielbeins, in terms of which the potential is polynomial. Doing so trades off the aforementioned complicated matrix equations for arguably simpler algebraic constraint equations\footnote{Unfortunately, the necessity of solving \emph{some} set of constraint equations, whether they be matrix equations or algebraic ones, seems to be inevitable. Indeed, we will find the same to be true for the procedure we are going to develop in Section \ref{sec:perts}.}; importantly, this happens \emph{without} affecting the kinetic structure, so both challenges associated to the initial approach from \cite{MG_arbitrary_BG,Syzygies,linear_spin2_gen_BG} are nicely bypassed. However, the approach of \cite{Volkov_arbitraryBG,Volkov_arbitraryBG_detailed} comes at the cost of having to parametrise the spin-2 field not with a genuine symmetric metric perturbation $h_{\mu\nu}$, but instead with a non-symmetric tensor $X_{\mu\nu}$ containing a maximal 16 independent components (in 4 dimensions). To show that the theory still propagates the correct number of degrees of freedom for a massive spin-2 field, one is forced to do lots of complex calculations to first compute the mass term in terms of this non-symmetric tensor, then to demonstrate that the equations of motion lead to just the right number of constraints to eliminate the unphysical degrees of freedom. The authors succeed in doing so, to their credit; it is just that the calculation is very complicated and as a result the underlying structure of the theory appears somewhat esoteric -- see appendix \ref{app:old perts} for the details. Again, it is worth stressing that this is still all at only the linear level, and that it does not easily generalise to additional interacting metrics or more dimensions.

The culmination of all this discussion is that we currently have three procedures for linearising multi-gravity theories around generic backgrounds, all of which are workable, but there is no one procedure that is totally satisfactory. We would like to improve this situation.

In this work, we are going to develop a new procedure to determine the perturbations of generic multi-gravity theories around arbitrary backgrounds. It is inspired by the procedure of \cite{Volkov_arbitraryBG,Volkov_arbitraryBG_detailed}, as we too start initially from the vielbein formalism and convert to the metric formalism later to avoid the complications of dealing with the matrix square roots. The key difference that distinguishes our approach from theirs, which is central to its utility, lies in how we relate the vielbein perturbations to the metric perturbations, and in how we compute the mass terms. We will see that our procedure has four main advantages over the previous approaches. Firstly, we argue that it is the simplest approach to date: the vielbein perturbations are split up explicitly into the physical degrees of freedom corresponding to the metric perturbations and unphysical degrees of freedom corresponding to local Lorentz transformations, avoiding use of the aforementioned non-symmetric tensor $X_{\mu\nu}$, thereby facilitating the expression of the spin-2 mass terms directly in terms of the metric perturbations from the outset. Secondly, it readily generalises to higher orders in perturbation theory. Thirdly, it makes abundantly clear the fact that the theory is ghost free, as it becomes obvious that $h_{00}$ will always survive as a Lagrange multiplier in the action at all orders, irrespective of the background. Fourthly, it works in both arbitrary spacetime dimension and for any number of interacting metrics.

The structure of the paper is as follows: in section \ref{sec:multigrav}, we review the fundamentals of multi-gravity at the background level, outlining both its metric and vielbein formulations and explaining how to relate them (and when it is possible to do so); in section \ref{sec:perts}, we develop our perturbation procedure and use it to derive the linearised field equations for multi-gravity theories around generic backgrounds; in section \ref{sec:examples}, we apply our formalism to recover and extend some results regarding the linearised field equations of massive gravity theories on proportional, cosmological and black hole backgrounds; in section \ref{sec:cubic}, we compute, for the first time, the cubic order multi-gravity potential around a generic background; finally, we conclude in section \ref{sec:conclusion}.

We work in natural units $c=\hbar=1$ throughout, and always use a mostly-plus metric signature.

\vspace{-0.7em}
\section{Review of multi-gravity}\label{sec:multigrav}

As discussed in the introduction, multi-gravity has two distinct formulations: one is known as the \emph{metric formalism}, where the interaction potential coupling the various metrics is built from those metrics directly, and the other is known as the \emph{vielbein formalism}, where it is instead built from wedge products of the various tetrad 1-forms associated to each of the metrics. The two formalisms are only equivalent if an important relation known as the Deser-van Nieuwenhuisen (DvN) symmetric vielbein condition holds, as we will see in the coming section.

\subsection{Metric formalism}\label{sec:metric}

In the metric formalism, the multi-gravity action for $N$ metrics interacting on a $D$-dimensional spacetime manifold $\mathcal{M}_D$ reads as follows:
\begin{align}
    I &= I_K + I_V + I_M[g_{(i)}]\label{MultigravAction}
    \\
    I_K &= \sum_{i=0}^{N-1} \frac{M_i^{D-2}}{2} \int \dd[D]x\, \sqrt{-\det g_{(i)}} R_{(i)}
    \\
    I_V &=  -\sum_{i,j}\int \dd[D]x\, \sqrt{-\det g_{(i)}} \sum_{m=0}^{D} \beta_m^{(i,j)} e_m(S_{i\rightarrow j}) \label{MultigravPotentialMetric} \; .
\end{align}
Each metric $\gi{i}$ gets its own Einstein-Hilbert kinetic term, and the ghost-free dRGT potential is built by summing up the elementary symmetric polynomials, $e_m$, of the building-block matrices $S_{i\rightarrow j}$:
\begin{align}
    \Sij{i}{j} &= \sqrt{g_{(i)}^{-1} g_{(j)}} \; ,\label{Sij}
    \\
    e_m(S) &= \frac{1}{m!}\gdelta{m}S^{\nu_1}_{\;\mu_1}\hdots S^{\nu_m}_{\;\mu_m} \; ,\label{sym pols}
\end{align}
together with some constant coefficients $\beta_m^{(i,j)}=\beta_m^{(j,i)}$ (of mass dimension $D$) to characterise the interactions between $\gi{i}$ and $\gi{j}$. In Eqs. \eqref{Sij} and \eqref{sym pols}, $S_{i\rightarrow j}$ is the square root matrix discussed at length in the introduction, defined in the sense that $(S^2_{i\rightarrow j})^\mu_{\;\nu}=g^{(i)\mu\lambda} g^{(j)}_{\lambda\nu}$, and $\gdelta{m}$ is the \emph{generalised Kronecker delta}, defined by antisymmetrising the product of $m$ standard Kronecker deltas:
\begin{equation}\label{gen_delta}
    \gdelta{m} = m!\delta^{\mu_1}_{[\nu_1}\hdots \delta^{\mu_m}_{\nu_m]} = \epsilon^{\mu_1\hdots\mu_m}\epsilon_{\nu_1\hdots\nu_m} \; ,
\end{equation}
where $\epsilon_{\mu_1\hdots\mu_m}$ is the totally antisymmetric Levi-Civita \emph{symbol} (i.e. tensor density). The antisymmetry properties of the generalised delta are wholly responsible for the ghost freedom of the dRGT interaction structure, as we will see more explicitly in section \ref{sec:perts}. For now, note that the FP mass term from Eq. \eqref{FP}, with $A=-1$, is nothing more than $\mathcal{L}_{\text{FP}}\sim\delta^{\mu\nu}_{\alpha\beta}h^\alpha_{\;\mu}h^\beta_{\;\nu}$. A nice property that will become very useful later on is:
\begin{equation}\label{ProductDeltas}
    \delta^{\mu_1\hdots\mu_p}_{\nu_1\hdots\nu_p}\delta^{\nu_1}_{\mu_1}\hdots\delta^{\nu_m}_{\mu_m} = \frac{(D-p+m)!}{(D-p)!} \delta^{\mu_{m+1}\hdots\mu_p}_{\nu_{m+1}\hdots\nu_p} \; .
\end{equation}

Regarding the $S_{i\rightarrow j}$ matrices, there is also some more to say, because in general the matrix square root is not a uniquely defined function. It transpires that the matrix square root that defines $\Sij{i}{j}$ from $S^2_{i\rightarrow j}=g_{(i)}^{-1}g_{(j)}$ must be the \emph{principal} root \cite{sqrtMatrix,matrix_analysis_book}, so that the action is guaranteed to be real. Furthermore, the light cones of $\gi{i}$ and $\gi{j}$ must intersect such that the two metrics share common timelike/spacelike directions, so that $\Sij{i}{j}$ transforms as a (1,1)-tensor under (diagonal) diffeomorphisms \cite{lightcones_theorem_bigravity}. These two conditions imply a number of additional useful properties, in particular:
\begin{enumerate}[label=\textit{(\roman*)}]
    \itemsep0em

    \item Swapping $i$ and $j$ inverts the matrix, $\Sij{i}{j}=\Sij{j}{i}^{-1}$.\label{Sij_property1}

    \item $S_{i\rightarrow j}$ and $S_{j\rightarrow i}$ are equivalent upon lowering an index with the appropriate metric, $g^{(i)}_{\mu\lambda} (S_{i\rightarrow j})^\lambda_{\;\nu} = g^{(j)}_{\mu\lambda} (S_{j\rightarrow i})^\lambda_{\;\nu}\equiv (S_{ij})_{\mu\nu}$\footnote{To see this, start from $S_{i\rightarrow j}^2=g_{(i)}^{-1}g_{(j)}$, then multiply by $g_{(i)}$ from the left and $S_{j\rightarrow i}=S_{i\rightarrow j}^{-1}$ from the right.}.\label{Sij_property2}

    \item The tensor $(S_{ij})_{\mu\nu}$ with both indices downstairs is symmetric, $(S_{ij})_{\mu\nu}=(S_{ij})_{\nu\mu}$.\label{Sij_property3}
\end{enumerate}

\begin{figure}[ht!]
\centering
    \includegraphics[width=0.48\textwidth]{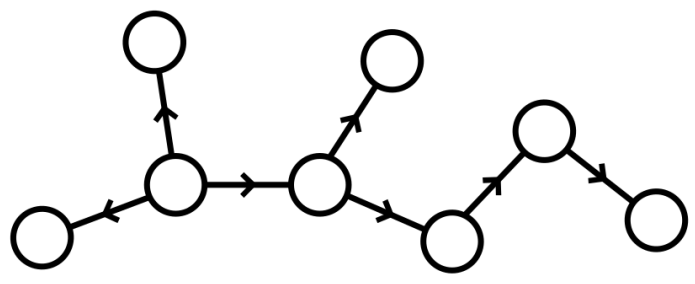}
    \caption{Directed theory graph representing some generic multi-metric theory. The circular nodes represent different metrics, the edges indicate interactions and the arrows point in the direction of positive interaction orientation. Each metric generically has a number of interactions of either orientation, and each edge contributes a term to the field equations of the two metrics it connects; these terms are orientation-dependent.}
    \label{fig:interaction structure}
\end{figure}

The simplest way to view the interaction structure of a given multi-metric theory is as a directed graph \cite{SC_and_graph_structure,prop_bg_multigrav}, as in figure \ref{fig:interaction structure}; the nodes correspond to metrics and the edges correspond to interactions. 

Interactions carry a sense of orientation thanks to property \ref{Sij_property1}: we say that a term in the potential, Eq. \eqref{MultigravPotentialMetric}, that explicitly contains $S_{i\rightarrow j}$ (\emph{not} $S_{j\rightarrow i}$) is positively oriented with respect to the $i$-th metric and negatively oriented with respect to the $j$-th metric. The orientation of an interaction with respect to a given metric affects the form of that metric's field equations, as we will soon see. It is also simply an artifact of the way one chooses to write down the potential and its interaction coefficients: the following identity holds on the building-blocks of the potential:
\begin{equation}\label{swap_orientation}
\begin{split}
    &\sqrt{-\det g_{(i)}}\sum_{m=0}^D \beta_m^{(i,j)}e_m(S_{i\rightarrow j}) 
    \\
    = &\sqrt{-\det g_{(j)}}\sum_{m=0}^D \beta_{D-m}^{(i,j)}e_m(S_{j\rightarrow i}) \; ,
\end{split}
\end{equation}
which shows that one can always consider any given positively oriented interaction as a negatively oriented one simply by redefining the interaction coefficients. An important final point is that multi-metric interactions are ghost-free so long as there are no cycles present (a cycle is e.g. $1\rightarrow2\rightarrow3\rightarrow1$, so that the potential is built from \emph{all three} of $S_{1\rightarrow2}$, $S_{2\rightarrow3}$ and $S_{3\rightarrow1}$; in other words, it is a loop in the theory graph) \cite{cycles,ghost_freedom_multigravity,3D_vbein_nocycles}.

The field equations arising from the action \eqref{MultigravAction} read as follows:
\begin{equation}\label{Einstein eqs}
    M^{D-2}_{i} \Gi{i} + \Wi{i} = \Tmunui{i} \; ,
\end{equation}
where the new term $W$ characterises the effect of the interactions over and above the standard GR interactions. It is given by:
\begin{equation}\label{W_metric}
    \Wi{i} = \sum_j [W_{i,j}^{(+)}]^\mu_{\;\nu}
    +\sum_k [W_{i,k}^{(-)}]^\mu_{\;\nu} \; ,
\end{equation}
where (with respect to the $i$-th metric) $j$ denote positively oriented interactions, $k$ denote negatively oriented interactions, and we define\footnote{A comment is in order here: typically, in most massive gravity literature, the $W$-tensors are expressed in terms of a matrix $Y_{(m)}(S)=\sum_{n=0}^m(-1)^n S^{m-n}e_n(S)$ as (e.g. for the positively oriented contributions) $W_{i,j}^{(+)} = \sum_{m=0}^D (-1)^m \beta_m^{(i,j)} Y_{(m)}(\Sij{i}{j})$; one can check that $Y_{(m)\nu}^\mu(S)=\frac{(-1)^m}{m!}\delta^{\mu\lambda_1\hdots\lambda_{m}}_{\nu\gamma_1\hdots\gamma_{m}}S^{\gamma_1}_{\;\lambda_1}\hdots S^{\gamma_{m}}_{\;\lambda_{m}}$, so this definition coincides with ours.\label{footnote_matrix}}:
\begin{align}
    [W_{i,j}^{(+)}]^\mu_{\;\nu} &= \sum_{m=0}^D \frac{\beta_m^{(i,j)}}{m!}\delta^{\mu\lambda_1\hdots\lambda_{m}}_{\nu\gamma_1\hdots\gamma_{m}}(\Sij{i}{j})^{\gamma_1}_{\;\lambda_1}\hdots (\Sij{i}{j})^{\gamma_{m}}_{\;\lambda_{m}} \; ,\label{Wp}
    \\
    [W_{i,k}^{(-)}]^\mu_{\;\nu} &= \sum_{m=0}^D \frac{\beta_{D-m}^{(k,i)}}{m!}\delta^{\mu\lambda_1\hdots\lambda_{m}}_{\nu\gamma_1\hdots\gamma_{m}}(\Sij{i}{k})^{\gamma_1}_{\;\lambda_1}\hdots (\Sij{i}{k})^{\gamma_{m}}_{\;\lambda_{m}}  \; , \label{Wm}
\end{align}
as the respective contributions from either orientation, whose indices are raised and lowered using the metric corresponding to the first of the two subscript Latin indices, which in this case is $\gi{i}$. The structure is the same for both contributions; the only real difference is that positively oriented interactions contribute with $\beta_m^{(i,j)}$, while negatively oriented interactions contribute with $\beta_{D-m}^{(k,i)}$. One can show that the two contributions satisfy the following algebraic identity \cite{EinsteinIsEinsteinBigravity}:
\begin{equation}\label{AlgebraicConstraint}
\begin{split}
    &\sqrt{-\det g_{(i)}}\, [W_{i,j}^{(+)}]^\mu_{\;\nu} + \sqrt{-\det g_{(j)}}\,[W_{j,i}^{(-)}]^\mu_{\;\nu} 
    \\
    &= \sqrt{-\det g_{(i)}}\,\delta^\mu_\nu\sum_{m=0}^{D} \beta_m^{(i,j)} e_m(S_{i\rightarrow j}) \; ,
\end{split}
\end{equation}
and thanks to property \ref{Sij_property3} of the $S_{i\rightarrow j}$ matrices, the $W$-tensors with both indices downstairs are symmetric, $W^{(i)}_{[\mu\nu]}=0$ (as they must be, given that the Einstein tensors and energy-momentum tensors are also symmetric).

Finally, the Bianchi identities on the Einstein tensors as well as invariance of the matter action $I_M$ under diagonal diffeomorphisms implies the following condition on the $W$-tensors, referred to as the \emph{Bianchi constraint}:
\begin{equation}\label{W sum}
    \sum_{i=0}^{N-1} \sqrt{-\det g_{(i)}}\nabla^{(i)}_{\mu}\Wi{i} = 0 \; .
\end{equation}
If matter couples only to one distinguished metric, or if there is no matter coupling at all (i.e. when one is in vacuum), then this condition is strengthened to:
\begin{equation}\label{Bianchi constraint}
    \nabla^{(i)}_{\mu}\Wi{i} = 0 \; ,
\end{equation}
informing us that there can be no flow of energy-momentum across the interactions between metrics.

\subsection{Vielbein formalism}\label{sec:vielbein}

In the vielbein formalism, one instead expresses everything given above in the convenient language of differential forms, using the tetrad 1-forms $e^{(i)a}=\vbein{i}{\mu}{a} \dd x^\mu$ in place of the metrics, with the vielbeins defined in the usual way through:
\begin{equation}\label{vbein def}
    \gi{i} = \vbein{i}{\mu}{a} \vbein{i}{\nu}{b} \eta_{ab} \; .
\end{equation}
The analogue of the multi-metric action \eqref{MultigravAction} is given by (see e.g. \cite{interacting_spin2,ClockworkCosmo,My_deconstruction}):
\begin{align}\label{MultigravActionVbein}
    I &= I_K + I_V + I_M[e^{(i)}]
    \\
    I_K &= \sum_{i=0}^{N-1} \frac{M_i^{D-2}}{2} \int_{\mathcal{M}_D} R^{(i)}_{ab} \wedge \hodge^{(i)} e^{(i)ab}\label{MultigravKinetic}
    \\
    I_V &= -\sum_{j_1\hdots j_D=0}^{N-1} \int_{\mathcal{M}_D} \varepsilon_{a_1\hdots a_D} T_{j_1\hdots j_D} e^{(j_1)a_1}\wedge\hdots\wedge e^{(j_D)a_D}\label{MultigravPotential}
\end{align}
where the kinetic term is just a rewriting of the standard Einstein-Hilbert term in terms of the curvature 2-form $R_{ab}=\frac12 R_{abcd}e^{cd}$, using the shorthand $e^{ab\hdots}\equiv e^a\wedge e^b\wedge\hdots$, and the potential is now built from the wedge products of the various tetrads, with some new symmetric coefficients $T_{j_1\hdots j_D}=T_{(j_1\hdots j_D)}$ (again of mass dimension $D$) to characterise the interactions.

These coefficients are analogous to the $\beta_m^{(i,j)}$ of the metric formalism, but they are not necessarily equivalent; indeed, as alluded to at the start of this section, not all multi-gravity theories described by the multi-vielbein action \eqref{MultigravActionVbein} can be equivalently expressed in the multi-metric language of Eq. \eqref{MultigravAction}. In fact, this happens only when the DvN condition:
\begin{equation}\label{SymmetricVierbeinCondition}
    \eta_{ab}\vbein{i}{[\mu}{a}\vbein{j}{\nu]}{b} = 0 \; ,
\end{equation}
is satisfied, which allows one to trade off products of vielbeins for the $\Sij{i}{j}$ matrices of the metric formalism. To understand why this is the case, and how it works in practice, it helps to first have a short discussion regarding the degree of freedom count in multi-gravity theories, which will also prove to be important in section \ref{sec:perts} when we develop our formalism for studying perturbations.

To begin, note that a generic $D$-dimensional vielbein is an invertible matrix without any symmetry, so contains $D^2$ independent components. The metric to which it is associated via Eq. \eqref{vbein def} is symmetric, containing only $\frac12 D(D+1)$ components; the residual $\frac12D(D-1)$ may be parametrised by a local Lorentz transformation on the vielbein, $\vbein{i}{\mu}{a}\rightarrow\tensor{\Lambda}{^{(i)a}_b}\vbein{i}{\mu}{b}$, as such transformations leave the metric invariant (since $\Lambda^T\eta\Lambda=\eta$). In the multi-vielbein action \eqref{MultigravActionVbein}, the kinetic terms and matter couplings are fully Lorentz invariant, so all dependence on $\tensor{\Lambda}{^{(i)a}_b}$ drops out of both $I_K$ and $I_M$. However, the interaction potential breaks full Lorentz invariance to just the diagonal subgroup transforming all vielbeins in the same way, hence the Lorentz fields remain present in $I_V$. They are thus auxiliary fields, which can (in principle) be eliminated in favour of the metric degrees of freedom.

To be more precise, one may always parametrise a generic vielbein as:
\begin{equation}
    \vbein{i}{\mu}{a} = \tensor{\Lambda}{^{(i)a}_b} E^{(i)b}_\mu \; ,
\end{equation}
where $\tensor{\Lambda}{^{(i)a}_b}$ are the local Lorentz matrices and $E^{(i)a}_\mu$ is a restricted vielbein containing only the metric degrees of freedom. The $\tensor{\Lambda}{^{(i)a}_b}$ matrices may be expressed in terms of another antisymmetric matrix $\omega^{(i)}_{ab}=-\omega^{(i)}_{ba}$, whose $\frac12D(D-1)$ components are the aforementioned auxiliary Lorentz fields, via the Cayley transform \cite{mass_spectrum_multivielbein}:
\begin{equation}
    \tensor{\Lambda}{^{(i)a}_b} = [(\eta+\omega^{(i)})^{-1}]^{ac}[\eta-\omega^{(i)}]_{bc} \; .
\end{equation}
The equations of motion for these Lorentz fields, $\delta I_V/\delta \omega^{(i)}_{ab}=0$, give the \emph{Lorentz constraints}, which one can show are entirely equivalent to the antisymmetric part of the vielbein field equations, $W^{(i)}_{[\mu\nu]}=0$, since \cite{metric_multigravity,Ghost_freedom_Multivielbein}:
\begin{equation}
    2\frac{\delta I_V}{\ivbein{i}{[\mu}{a}}\eta_{ab}\vbein{i}{\nu]}{b} = [\eta+\omega^{(i)}]_{ac}\vbein{i}{\mu}{c}\frac{\delta I_V}{\delta \omega^{(i)}_{ab}}[\eta+\omega^{(i)}]_{bd}\vbein{i}{\nu}{d} \; .
\end{equation}

Explicitly, the $W$-tensors in vielbein form read as follows \cite{BHs_multigrav}:
\begin{equation}\label{W tensor vielbein}
\begin{split}
    \Wi{i} &= e^{(i)\mu\lambda_2\hdots\lambda_D}_{\;\;\;\;a b_2\hdots b_D} \\
    &\times\sum_{j_2\hdots j_{D}} \mathcal{P}(i) T_{ij_2\hdots j_{D}} \vbein{i}{\nu}{a} \vbein{j_2}{\lambda_2}{b_2}\hdots\vbein{j_{D}}{\lambda_{D}}{b_{D}} \; ,
\end{split}
\end{equation}
defining the mixed index version of the generalised Kronecker delta:
\begin{equation}\label{VielbeinDeltas}
    e^{(i)\mu_1\hdots\mu_D}_{\;\;\;\;a_1\hdots a_D} = \ivbein{i}{\nu_1}{a_1}\hdots\ivbein{i}{\nu_D}{a_D}\delta^{\mu_1\hdots\mu_D}_{\nu_1\hdots\nu_D}\; ,
\end{equation}
and where $\mathcal{P}(i)$ counts the number of times the index $i$ appears in the interaction coefficients (i.e. a term with $T_{ij_2\hdots j_{D}}$ has $\mathcal{P}(i)=1$, a term with $T_{iij_3\hdots j_{D}}$ has $\mathcal{P}(i)=2$, and so on). Enforcing $W^{(i)}_{[\mu\nu]}=0$ therefore leads to the following set of $N\times\frac12D(D-1)$ algebraic equations:
\begin{equation}\label{LorentzConstraintsNonlin}
    g^{(i)}_{\lambda[\mu}u^{(i)\lambda}_{\;\;\;\;\;a}\vbein{i}{\nu]}{a} = 0 \implies \eta_{ab}u_{[\mu}^{(i)a} e^{(i)b}_{\nu]} = 0 \; ,
\end{equation}
where we have defined:
\begin{equation}
    u^{(i)\mu}_{\;\;\;\;\;a} = e^{(i)\mu\lambda_2\hdots\lambda_D}_{\;\;\;\;a b_2\hdots b_D}\sum_{j_2\hdots j_{D}} \mathcal{P}(i) T_{ij_2\hdots j_{D}} \vbein{j_2}{\lambda_2}{b_2}\hdots\vbein{j_{D}}{\lambda_{D}}{b_{D}} \; ;
\end{equation}
Eqs. \eqref{LorentzConstraintsNonlin} are the Lorentz constraints. These constraints are not all linearly independent: the weighted sum corresponding to $\eta_{ab}u^{(i)a}_{[\mu}u^{(i)b}_{\nu]}$ clearly vanishes, so they actually only eliminate $(N-1)\times\frac12D(D-1)$ degrees of freedom. A further $\frac12D(D-1)$ must be eliminated in some other way to ensure that only the metric degrees of freedom remain dynamical; this is the role played by the single surviving diagonal copy of Lorentz invariance, which ensures that precisely this many local Lorentz fields drop out of the action upon fixing a gauge.

In general, solutions to Eqs. \eqref{LorentzConstraintsNonlin} are not known, so as of yet most multi-vielbein theories have no known metric formulation\footnote{It is also worth noting that not all multi-vielbein theories are ghost-free -- it seems that only a small subset of them remain so, namely: those with pairwise interactions permitting a clean metric formalism as we will describe, and those in which the $\Tcoeffs$ factorise as $T_{j_1}\hdots T_{j_D}$ (whose metric formulation is not yet known) \cite{vielbein_to_rescue,beyond_pairwise_couplings,Ghost_freedom_Multivielbein}.}. However, for theories exhibiting exclusively pairwise interactions between neighbouring vielbeins, where the $\Tcoeffs$ are restricted to only terms of the form $T_{iiii\hdots}$, $T_{jiii\hdots}$, $T_{jjii\hdots}$ etc., Eq. \eqref{LorentzConstraintsNonlin} reduces to the DvN condition \eqref{SymmetricVierbeinCondition} for each pair of interacting vielbeins \cite{vielbein_to_rescue}. As we stated earlier, whenever the DvN condition holds, the multi-vielbein theory described above becomes equivalent to the multi-metric theory described in the previous section. To see how this works, note that in matrix notation the DvN condition reads ${e_{(i)}^T\eta e_{(j)}=e_{(j)}^T\eta e_{(i)}}$, from which one finds $e_{(j)}e_{(i)}^{-1}=\eta^{-1}(e_{(i)}^T)^{-1}e_{(j)}^T\eta$, and hence:
\begin{align}
    S^2_{i\rightarrow j} &= g_{(i)}^{-1}g_{(j)}\nonumber
    \\
    &= \left(e_{(i)}^T\eta e_{(i)}\right)^{-1}\left(e_{(j)}^T\eta e_{(j)}\right)\nonumber
    \\
    &= e_{(i)}^{-1}\eta^{-1}(e_{(i)}^T)^{-1}e_{(j)}^T\eta e_{(j)}\nonumber
    \\
    &= \left(e_{(i)}^{-1}e_{(j)}\right)^2\; .
\end{align}
Taking the principal square root, one arrives at:
\begin{equation}\label{SijVielbein}
    (S_{i\rightarrow j})^\mu_{\;\nu} = \ivbein{i}{\mu}{a}\vbein{j}{\nu}{a} \; .
\end{equation}
The DvN condition is thus equivalent to the statement from the metric formalism that $(S_{ij})_{\mu\nu}=(S_{ij})_{\nu\mu}$, which was likewise necessary to ensure $W^{(i)}_{[\mu\nu]}=0$.

Armed with a means of relating the vielbeins to the metrics for pairwise interactions, one may now consider the subset of the multi-vielbein potential \eqref{MultigravPotential} for which the DvN condition is known to apply i.e. interactions that couple together $m$ $e^{(j)}$'s and ${(D-m)}$ $e^{(i)}$'s. Such terms are characterised by interaction coefficients that are of the form $T_{\{j\}^m\{i\}^{D-m}}$, and owing to the symmetry of these coefficients, there are $\binom{D}{m}$ such terms in total. Expanding this particular subset of the potential out into components, one gets \cite{interacting_spin2}:
\begin{widetext}
\begin{equation}\label{MetricVbeinEquivalence}
    \begin{split}
        I_V &\supset -\sum_{i,j} \int_{\mathcal{M}_D}T_{\{j\}^m\{i\}^{D-m}}\varepsilon_{a_1\hdots a_D} e^{(j)a_1}\wedge\hdots\wedge e^{(j)a_m}\wedge e^{(i)a_{m+1}}\wedge\hdots\wedge e^{(i)a_D}
        \\
        &= -\int \dd[D]x \sqrt{-\det g_{(i)}}\,\binom{D}{m}T_{\{j\}^m\{i\}^{D-m}}e^{(i)\mu_1\hdots\mu_D}_{\;\;\;\;a_1\hdots a_D} \vbein{j}{\mu_1}{a_1}\hdots\vbein{j}{\mu_m}{a_m}\vbein{i}{\mu_{m+1}}{a_{m+1}}\hdots\vbein{i}{\mu_D}{a_D}
        \\
        &= -\int \dd[D]x \sqrt{-\det g_{(i)}} \,(D-m)!\binom{D}{m}T_{\{j\}^m\{i\}^{D-m}} e^{(i)\mu_1\hdots\mu_m}_{\;\;\;\;a_1\hdots a_m} \vbein{j}{\mu_1}{a_1}\hdots\vbein{j}{\mu_m}{a_m}
        \\
        &= -\int \dd[D]x \sqrt{-\det g_{(i)}} \, D!\,T_{\{j\}^m\{i\}^{D-m}} \frac{1}{m!} \delta^{\mu_1\hdots \mu_m}_{\nu_1\hdots\nu_m} (S_{i\rightarrow j})^{\nu_1}_{\;\mu_1}\hdots(S_{i\rightarrow j})^{\nu_m}_{\;\mu_m}
        \\
        &= -\int \dd[D]x \sqrt{-\det g_{(i)}}\, D!\,T_{\{j\}^m\{i\}^{D-m}} e_m(S_{i\rightarrow j}) \; ,
    \end{split}
\end{equation}
\end{widetext}
where on the third line we used Eq. \eqref{ProductDeltas}, on the fourth line we used Eq. \eqref{SijVielbein}, and on the final line we used Eq. \eqref{sym pols}. The full multi-vielbein potential \eqref{MultigravPotential} is then obtained by summing up all the different contributions of the type \eqref{MetricVbeinEquivalence}, coming from every possible value of $m$, for all combinations of $i$ and $j$, i.e.
\begin{equation}
\begin{split}
    &I_V =\\
    &-\sum_{i,j}\int \dd[D]x \sqrt{-\det g_{(i)}} \sum_{m=0}^D D!\, T_{\{j\}^m\{i\}^{D-m}} e_m(S_{i\rightarrow j}) \; .
\end{split}
\end{equation}
Comparing against the multi-metric potential \eqref{MultigravPotentialMetric}, one sees that, for the subclass of pairwise interactions where the DvN condition holds, the $\Tcoeffs$ of the vielbein formalism are related to the $\beta_m^{(i,j)}$ of the metric formalism by:
\begin{align}
    D!\,T_{iiii\hdots i} &= \sum_j\beta_0^{(i,j)} + \sum_k\beta_D^{(k,i)}\label{betas1} \\
    D!\,T_{\{j\}^m\{i\}^{D-m}}  &= \beta_m^{(i,j)}\label{betas2} \; ,
\end{align}
where, as in section \ref{sec:metric}, $j$ and $k$ refer respectively to positively and negatively oriented interactions with respect to $\gi{i}$. The sense of interaction orientation from the metric formalism is hence encoded in the vielbein formalism within the structure of $T_{iiii\hdots i}$.

In a similar vein, one may show that Eq. \eqref{W tensor vielbein} for the vielbein $W$-tensor is equivalent to Eq. \eqref{W_metric} from the metric formalism whenever the interactions are strictly pairwise, by substituting in Eqs. \eqref{betas1} and \eqref{betas2} for the $\Tcoeffs$, then using Eqs. \eqref{ProductDeltas} and \eqref{SijVielbein} to rewrite everything in terms of the building-block matrices $\Sij{i}{j}$; one eventually identifies the terms appearing in Eqs. \eqref{Wp} and \eqref{Wm}.

\section{Multi-gravity perturbations: general formalism}\label{sec:perts}

We now possess the necessary technology required to develop our formalism for studying perturbations. We begin with the multi-vielbein action \eqref{MultigravActionVbein}, and perturb the vielbeins directly, $\vbein{i}{\mu}{a}=\bar{e}^{(i)a}_\mu+\delta\vbein{i}{\mu}{a}$. We would like to eventually relate the vielbein perturbations to the corresponding metric perturbations and the $\Sij{i}{j}$ matrices; any such relation must also account for the Lorentz constraints, as per the discussion in the previous section. Thankfully, an all-orders expansion of a generic vielbein in terms of the metric perturbations, $\dg{i}{\mu}{\nu}$, and local Lorentz perturbations $\om{i}{\mu}{\nu}=\bar{e}^{(i)\mu}_{\;\;\;\;\;a}\tensor{\omega}{^{(i)a}_b}\bar{e}^{(i)b}_\nu$, has already been developed in \cite{mass_spectrum_multivielbein}; it reads as follows\footnote{If one wishes to compare against the vielbein expansion used in \cite{Volkov_arbitraryBG,Volkov_arbitraryBG_detailed}, one should note that the non-symmetric tensor $X_{\mu\nu}$ they use to parametrise their spin-2 field, which we mentioned in the introduction, is defined by $X^\mu_{\;\nu}\equiv\bar{e}^\mu_{\;a}\delta e^{\;a}_\nu$; our expression \eqref{VielbeinPerturbations} simply divides this tensor explicitly into the components coming from the metric perturbations, which are physical, and the components coming from the local Lorentz perturbations, which are unphysical and will be later eliminated by constraints. We discuss this in more detail in appendix \ref{app:Volkov}.\label{X_footnote}}:
\begin{align}
    \vbein{i}{\mu}{a} &= \bar{e}^{(i)a}_\mu + \bar{e}^{(i)a}_\nu \sum_{n=1}^\infty \kappa_i^n \Bigg[\binom{\frac12}{n}(\delta g_{(i)}^n)^\nu_{\;\mu} \nonumber \\
    &\qquad\qquad\qquad+2\sum_{m=1}^n\binom{\frac12}{n-m}(\omega_{(i)}^m)^\nu_{\;\lambda}(\delta g_{(i)}^{n-m})^\lambda_{\;\mu}\Bigg]\; ,\label{VielbeinPerturbations}
\end{align}
where the fractional binomial coefficient is defined by:
\begin{equation}
    \binom{\frac12}{n} = 
    \begin{cases}
        1 & n=0
        \\
        \frac{(-1)^{n-1}}{2^{2n-1}n}\binom{2n-2}{n-1} & n>0
    \end{cases} \; ,
\end{equation}\vspace{0.8em}

\noindent and we include factors of $\kappa_i=1/M_i^{(D-2)/2}$ to ensure canonical normalisation of the metric perturbations. One should keep in mind that the local Lorentz fields $\om{i}{\mu}{\nu}$ are ultimately non-dynamical and should eventually be able to be eliminated in favour of $\dg{i}{\mu}{\nu}$ through the antisymmetric part of the perturbed field equations.

One may rewrite the expansion \eqref{VielbeinPerturbations} as a differential form expression, expanding the tetrad 1-forms in the following way:
\begin{equation}
    e^{(i)a} = \bar{e}^{(i)a} + \kappa_i \delta e^{(i)a}_{(1)} + \kappa_i^2 \delta e_{(2)}^{(i)a} + \kappa_i^3 \delta e_{(3)}^{(i)a} + \hdots \; ,\label{1formperts}
\end{equation}
where we identify the 1-form perturbations to each order as:
\begin{widetext}
\begin{align}
    \delta e^{(i)a}_{(1)} &= \frac12\bar{e}^{(i)a}_\nu \left[\dg{i}{\nu}{\mu}-4\om{i}{\nu}{\mu}\right] \dd x^\mu \; ,\label{1storder}
    \\
    \delta e^{(i)a}_{(2)} &= -\frac18\bar{e}^{(i)a}_\nu\bigg[\dg{i}{\nu}{\lambda}\dg{i}{\lambda}{\mu} + 8 \dg{i}{\nu}{\lambda}\om{i}{\lambda}{\mu} - 16\om{i}{\nu}{\lambda}\om{i}{\lambda}{\mu} \bigg]\dd x^\mu \; ,\label{2ndorder}
    \\
    \delta e_{(3)}^{(i)a} &= \frac{1}{16}\bar{e}_\nu^{(i)a} \bigg[\dg{i}{\nu}{\lambda}\dg{i}{\lambda}{\rho}\dg{i}{\rho}{\mu}-4\dg{i}{\nu}{\lambda}\dg{i}{\lambda}{\rho}\om{i}{\rho}{\mu}+16\dg{i}{\nu}{\lambda}\om{i}{\lambda}{\rho}\om{i}{\rho}{\mu}+32\om{i}{\nu}{\lambda}\om{i}{\lambda}{\rho}\om{i}{\rho}{\mu}\bigg]\dd x^\mu \; ,\label{3rdorder}
\end{align}
\end{widetext}
One may then simply substitute these expressions into the multi-vielbein action up to the desired order in perturbation theory. Let us work to quadratic order for now, so that we may determine the linearised multi-gravity field equations; we will compute the cubic order potential later on in section \ref{sec:cubic}. 

\subsection{Quadratic action and linearised field equations}

The second-order variation of the Einstein-Hilbert term is well-known and reads as follows:
\begin{align}
    I_K^{(2)} &= \sum_{i=0}^{N-1} \int \dd[D]x \sqrt{-\det \bar{g}_{(i)}}
    \Bigg[-\frac14 \delta g^{(i)\nu}_{\;\;\;\;\;\mu}\tensor{\bar{\mathcal{E}}}{^{(i)\mu}_\nu^\alpha_\beta}\delta g^{(i)\beta}_{\;\;\;\;\;\alpha}\nonumber
    \\
    &-\frac18 \delta g^{(i)\mu\nu}\left(\bar{g}^{(i)}_{\mu\nu}\bar{R}^{(i)\alpha\beta} - \bar{R}^{(i)}\delta^\alpha_\mu\delta^\beta_\nu\right)\delta g^{(i)}_{\alpha\beta} \nonumber
    \\
    &+ \frac12 \bar{G}^{(i)}_{\alpha\beta}\left(\delta g^{(i)\alpha}_{\;\;\;\;\;\lambda}\delta g^{(i)\lambda\beta} - \frac14 \delta g^{(i)}\delta g^{(i)\alpha\beta}\right)\Bigg] \label{Kinetic2ndOrder}
\end{align}
The first term in this expression is the standard FP kinetic term, where the curved spacetime \emph{Lichnerowicz operator} is given by:
\begin{equation}
    \tensor{\bar{\mathcal{E}}}{^{(i)\mu}_\nu^\alpha_\beta} 
    = \frac12\delta^{\mu\rho\alpha}_{\nu\sigma\beta}\bar{\nabla}^{(i)\sigma}\bar{\nabla}^{(i)}_\rho \; ,
\end{equation}
while the remaining two terms encode the explicit contributions to the action coming from the background curvature (which of course vanish around flat spacetime). As expected, the local Lorentz fields $\om{i}{\mu}{\nu}$ are not present in the kinetic term at all, since the Einstein-Hilbert action is Lorentz invariant.

The second-order variation of the potential term, using Eq. \eqref{1formperts} and including only the terms quadratic in the metric perturbations, has two contributions:
\begin{align}
    I_V^{(2)} = &-\sum_{ijk_3\hdots k_{D}} \int\kappa_i\kappa_j \binom{D}{2}T_{ijk_3\hdots k_{D}} \varepsilon_{abc_3\hdots c_{D}} \nonumber
    \\
    &\qquad\qquad\times\delta e^{(i)a}_{(1)}\wedge \delta e^{(j)b}_{(1)}\wedge \bar{e}^{(k_3)c_3}\wedge\hdots\wedge\bar{e}^{(k_{D})c_{D}} \nonumber
    \\
    &-\sum_{ik_2\hdots k_{D}}\int\kappa_i^2\binom{D}{1} T_{ik_2\hdots k_D}\varepsilon_{ac_2\hdots c_D}\nonumber
    \\
    &\qquad\qquad\times\delta e^{(i)a}_{(2)} \wedge \bar{e}^{(k_2)c_2}\wedge\hdots\wedge\bar{e}^{(k_D)c_D} \; ,
\end{align}
where we have exploited the symmetry of the $\Tcoeffs$ coefficients, as well as the antisymmetry of the $\varepsilon$-tensor and wedge products to bring all of the perturbation 1-forms to the front. Expanding out into components with Eqs. \eqref{1storder}--\eqref{3rdorder}, we explicitly have:
\begin{align}
    &I_V^{(2)} = -\sum_{ijk} \int\dd[D]x\sqrt{-\det \bar{g}_{(i)}} \nonumber
    \\
    &\times\Bigg[\frac{\kappa_i\kappa_j}{4} \binom{D}{2}T_{ijk_3\hdots k_{D}} \bar{e}^{(i)\mu\nu\lambda_3\hdots\lambda_D}_{\;\;\;\;abc_3\hdots c_D}\nonumber
    \\
    &\qquad\qquad\times(\delta g^{(i)}-4\omega^{(i)})^\alpha_{\;\mu}(\delta g^{(j)}-4\omega^{(j)})^\beta_{\;\nu}\nonumber
    \\
    &\qquad\qquad\times \bar{e}^{(i)a}_\alpha \bar{e}^{(j)b}_\beta \bar{e}^{(k_3)c_3}_{\lambda_3}\hdots\bar{e}^{(k_D)c_D}_{\lambda_D}\nonumber
    \\
    &\;\;\;\;\; -\frac{\kappa_i^2}{8} \binom{D}{1}T_{ik_2\hdots k_{D}} \bar{e}^{(i)\mu\lambda_2\hdots\lambda_D}_{\;\;\;\;ac_2\hdots c_D} \nonumber
    \\
    &\qquad\qquad\times\left(\delta g^{(i)\alpha}_{\;\;\;\;\;\beta}\delta g^{(i)\beta}_{\;\;\;\;\;\mu}-8g^{(i)\alpha}_{\;\;\;\;\;\beta}\omega^{(i)\beta}_{\;\;\;\;\;\mu}+16\omega^{(i)\alpha}_{\;\;\;\;\;\beta}\omega^{(i)\beta}_{\;\;\;\;\;\mu}\right)\nonumber
    \\
    &\qquad\qquad\times\bar{e}^{(i)a}_\alpha \bar{e}^{(k_2)c_2}_{\lambda_2}\hdots\bar{e}^{(k_D)c_D}_{\lambda_D}\Bigg] \; .
\end{align}
Now let us restrict to only pairwise interactions, so that the DvN condition \eqref{SymmetricVierbeinCondition} holds and the background vielbeins can be related to the $\bar{S}_{i\rightarrow j}$ matrices of the metric formalism through Eq. \eqref{SijVielbein}. Proceeding in a similar manner to how we did in Eq. \eqref{MetricVbeinEquivalence}, using the symmetry of the $\Tcoeffs$ coefficients, their relation to the $\beta_m^{(i,j)}$ of the metric formalism as given by Eqs. \eqref{betas1} and \eqref{betas2}, as well as Eqs. \eqref{ProductDeltas} and \eqref{VielbeinDeltas} to evaluate any products of generalised deltas arising, we find that the metric formalism version of the above expression reads as follows:
\begin{align}
        &I_V^{(2)} = \sum_{i=0}^{N-1} \int\dd[D]x\sqrt{-\det \bar{g}_{(i)}} \nonumber
        \\
        &\times\Bigg[ \frac{\kappa_i^2}{8} (\delta g^{(i)\nu}_{\;\;\;\;\;\sigma}\delta g^{(i)\sigma}_{\;\;\;\;\;\mu}-8\delta g^{(i)\nu}_{\;\;\;\;\;\sigma}\omega^{(i)\sigma}_{\;\;\;\;\;\mu}+16\omega^{(i)\nu}_{\;\;\;\;\;\sigma}\omega^{(i)\sigma}_{\;\;\;\;\;\mu})\bar{W}^{(i)\mu}_{\;\;\;\;\;\;\nu} \nonumber
        \\
        &-\frac{\kappa_i^2}{8}\tensor{\mathcal{A}}{^{(i)\mu}_{\alpha}^\nu_{\beta}}(\delta g^{(i)} - 4\omega^{(i)})^\alpha_{\;\mu}(\delta g^{(i)}-4\omega^{(i)})^\beta_{\;\nu} \nonumber
        \\
        & - \sum_j \frac{\kappa_i\kappa_j}{4} [\mathcal{B}_{i,j}^{(+)}]^{\mu\;\;\nu}_{\;\;\alpha\;\;\beta}(\bar{S}_{i\rightarrow j})^\beta_{\;\sigma} \nonumber
        \\
        &\qquad\times(\delta g^{(i)} - 4\omega^{(i)})^\alpha_{\;\mu}(\delta g^{(j)}-4\omega^{(j)})^\sigma_{\;\nu}\nonumber
        \\
        &- \sum_k \frac{\kappa_i\kappa_k}{4} [\mathcal{B}_{i,k}^{(-)}]^{\mu\;\;\nu}_{\;\;\alpha\;\;\beta}(\bar{S}_{i\rightarrow k})^\beta_{\;\sigma}\nonumber
        \\
        &\qquad\times(\delta g^{(i)} - 4\omega^{(i)})^\alpha_{\;\mu}(\delta g^{(k)}-4\omega^{(k)})^\sigma_{\;\nu}\Bigg] \; , \label{Potential2ndOrder}
\end{align}
where we recall that the background $W$-tensor is given by Eq. \eqref{W_metric}, and we have defined the following three new tensors:
\begin{align}
         &\tensor{\mathcal{A}}{^{(i)\mu}_{\alpha}^\nu_{\beta}} \nonumber
         \\
         &=\sum_j\sum_{m=0}^D\frac{\beta_m^{(i,j)}}{m!}\delta^{\mu\nu\lambda_1\hdots\lambda_m}_{\alpha\beta\gamma_1\hdots\gamma_m} (\bar{S}_{i\rightarrow j})^{\gamma_1}_{\;\lambda_1}\hdots(\bar{S}_{i\rightarrow j})^{\gamma_m}_{\;\lambda_m}\nonumber
         \\
         &\;+\sum_k\sum_{m=0}^D\frac{\beta_{D-m}^{(k,i)}}{m!}\delta^{\mu\nu\lambda_1\hdots\lambda_m}_{\alpha\beta\gamma_1\hdots\gamma_m} (\bar{S}_{i\rightarrow k})^{\gamma_1}_{\;\lambda_1}\hdots(\bar{S}_{i\rightarrow k})^{\gamma_m}_{\;\lambda_m}\label{Atens} \; ,
         \\
         &[\mathcal{B}_{i,j}^{(+)}]^{\mu\;\;\nu}_{\;\;\alpha\;\;\beta} \nonumber
         \\&=\sum_{m=0}^D\frac{\beta_m^{(i,j)}}{(m-1)!}\delta^{\mu\nu\lambda_1\hdots\lambda_{m-1}}_{\alpha\beta\gamma_1\hdots\gamma_{m-1}} 
         (\bar{S}_{i\rightarrow j})^{\gamma_1}_{\;\lambda_1}\hdots(\bar{S}_{i\rightarrow j})^{\gamma_{m-1}}_{\;\lambda_{m-1}}\label{Btens} \; ,
         \\
         &[\mathcal{B}_{i,k}^{(-)}]^{\mu\;\;\nu}_{\;\;\alpha\;\;\beta} \nonumber
         \\
         &=\sum_{m=0}^D\frac{\beta_{D-m}^{(k,i)}}{(m-1)!}\delta^{\mu\nu\lambda_1\hdots\lambda_{m-1}}_{\alpha\beta\gamma_1\hdots\gamma_{m-1}} (\bar{S}_{i\rightarrow k})^{\gamma_1}_{\;\lambda_1}\hdots(\bar{S}_{i\rightarrow k})^{\gamma_{m-1}}_{\;\lambda_{m-1}} \; .\label{Ctens}
\end{align}
It is possible to show that the $\mathcal{A}$, $\mathcal{B}$ and $\bar{W}$-tensors satisfy the following relationship when contracted with some generic tensor $\chi^\mu_{\;\nu}$ in a particular manner:
\begin{equation}\label{ABW_relation}
\begin{split}
    &\tensor{\mathcal{A}}{^{(i)\mu}_{\nu}^\alpha_{\beta}} \tensor{\chi}{^{\beta}_\alpha} 
    \\
    &+ \sum_j [\mathcal{B}_{i,j}^{(+)}]^{\mu\;\;\alpha}_{\;\;\nu\;\;\beta}\tensor{\chi}{^{\lambda}_\alpha}(\bar{S}_{i\rightarrow j})^\beta_{\;\lambda} 
    \\
    &+ \sum_k [\mathcal{B}_{i,k}^{(-)}]^{\mu\;\;\alpha}_{\;\;\nu\;\;\beta}\tensor{\chi}{^{\lambda}_\alpha}(\bar{S}_{i\rightarrow k})^\beta_{\;\lambda} = \delta^{\mu\alpha}_{\lambda\beta}\tensor{\chi}{^{\beta}_{\alpha}}\tensor{\bar{W}}{^{(i)\lambda}_\nu} \; .
\end{split}
\end{equation}
This will allow us to eliminate $\mathcal{A}$ from the action in favour of $\mathcal{B}$ and $\bar{W}$.

The linearised field equations for the metric perturbations now follow readily from the second-order action by varying Eqs. \eqref{Kinetic2ndOrder} and \eqref{Potential2ndOrder} with respect to $\delta g^{(i)\nu}_{\;\;\;\;\;\mu}$ and using Eq. \eqref{ABW_relation} to eliminate $\mathcal{A}$. They read as follows:
\begin{equation}\label{GeneralLinearisedFieldEqs}
    \begin{split}
        &\tensor{\bar{\mathcal{E}}}{^{(i)\mu}_\nu^\alpha_\beta}\tensor{\delta g}{^{(i)\beta}_\alpha} + \tensor{\mathcal{R}}{^{(i)\mu}_\nu^\alpha_\beta}\tensor{\delta g}{^{(i)\beta}_\alpha}
        \\
        &-\frac{\kappa_i^2}{2}\left[\tensor{\bar{W}}{^{(i)\mu}_\lambda}\tensor{\delta g}{^{(i)\lambda}_\nu} + \tensor{\bar{W}}{^{(i)\lambda}_\nu}\tensor{\delta g}{^{(i)\mu}_\lambda} -\tensor{\bar{W}}{^{(i)\mu}_\nu}\delta g^{(i)}\right]
        \\
        &+\frac{\kappa_i}{2}\sum_j[\mathcal{B}_{i,j}^{(+)}]^{\mu\;\;\alpha}_{\;\;\nu\;\;\beta}(\bar{S}_{i\rightarrow j})^\beta_{\;\lambda} \left\{[\tensor{\delta g}{^{\lambda}_\alpha}]_{i,j}-4[\tensor{\omega}{^\lambda_\alpha}]_{i,j}\right\}
        \\
        &+\frac{\kappa_i}{2}\sum_k[\mathcal{B}_{i,k}^{(-)}]^{\mu\;\;\alpha}_{\;\;\nu\;\;\beta}(\bar{S}_{i\rightarrow k})^\beta_{\;\lambda}\left\{[\tensor{\delta g}{^{\lambda}_\alpha}]_{i,k}-4[\tensor{\omega}{^\lambda_\alpha}]_{i,k}\right\}
        \\
        &= \kappa_i \Tmunui{i} \; ,
    \end{split}
\end{equation}
where the background curvature piece is:
\begin{equation}
\begin{split}
    &\tensor{\mathcal{R}}{^{(i)\mu}_\nu^\alpha_\beta}\tensor{\delta g}{^{(i)\beta}_\alpha} 
    \\
    &= \frac14\left[\delta^\mu_\nu \tensor{\bar{R}}{^{(i)\alpha}_\beta}\tensor{\delta g}{^{(i)\beta}_\alpha}+\tensor{\bar{R}}{^{(i)\mu}_\nu}\delta g^{(i)} - 2\bar{R}^{(i)}\tensor{\delta g}{^{(i)\mu}_\nu}\right]
    \\
    &-\bigg[\tensor{\bar{G}}{^{(i)\mu}_\lambda}\tensor{\delta g}{^{(i)\lambda}_\nu}+\tensor{\bar{G}}{^{(i)\lambda}_\nu}\tensor{\delta g}{^{(i)\mu}_\lambda}
    \\
    &\qquad-\frac14\left(\delta^\mu_\nu\tensor{\bar{G}}{^{(i)\alpha}_\beta}\tensor{\delta g}{^{(i)\beta}_\alpha}+ \tensor{\bar{G}}{^{(i)\mu}_\nu}\delta g^{(i)}\right)\bigg] \; ,
\end{split}
\end{equation}
and we have introduced the following notation for brevity:
\begin{equation}\label{DiffMetPerts}
    [x]_{i,j} = \kappa_j x_{(j)} - \kappa_i x_{(i)} \; .
\end{equation}

The first three terms in the linearised equations \eqref{GeneralLinearisedFieldEqs} are all related to the background curvature, and are essentially the same as they are in GR, if one thinks of the background $W$-tensor as an effective background source (in particular, around proportional solutions, it behaves like a cosmological constant, as we will see in section \ref{sec:examples}), while the final two are the terms genuinely arising due to the spin-2 interactions. This is reflected by the fact they are the only terms containing the Lorentz perturbations. Thus, the $\mathcal{B}$-tensors encode the effective masses\footnote{Here, we are defining an `effective mass' term as an interaction term appearing in the action that is quadratic in the metric perturbations. This is a slight abuse of language: in general, the background spacetime may not be an Einstein space, in which case the structure of the perturbations will not be Fierz-Pauli, so it is not clear that one can really identify these terms with the \emph{bona fide} spin-2 masses. Nevertheless, the nomenclature will suffice.} of the corresponding spin-2 fields on arbitrary backgrounds.

As discussed in section \ref{sec:vielbein}, the antisymmetric part of the field equations, after lowering an index with $\gi{i}$, should give the linearised Lorentz constraints. The only terms in Eq. \eqref{GeneralLinearisedFieldEqs} that are not manifestly symmetric are the $\mathcal{B}$-tensor terms; setting their antisymmetric part to zero results in the following set of algebraic constraints:
\begin{align}
    (S_{ij})_{\lambda[\mu}\left\{[\tensor{\delta g}{^{\lambda}_{\nu]}}]_{i,j}-4[\tensor{\omega}{^\lambda_{\nu]}}]_{i,j}\right\} = 0 \; ,
\end{align}
or, written more explicitly:
\begin{align}\label{Linearised_Lorentz_constraints}
    &\kappa_j (S_{ij})_{\mu\lambda}(\delta g^{(j)}-4\omega^{(j)})^\lambda_{\;\nu} + \kappa_i (S_{ij})_{\nu\lambda}(\delta g^{(i)}-4\omega^{(i)})^\lambda_{\;\mu} \nonumber
    \\
    &= [\mu\leftrightarrow\nu]\; ,
\end{align}
for every pair of interacting metrics. Eq. \eqref{Linearised_Lorentz_constraints} is of course consistent with the linearisation of the DvN condition \eqref{SymmetricVierbeinCondition}, using our vielbein expansion \eqref{VielbeinPerturbations}. It also agrees with the analogous vielbein expression derived in \cite{mass_spectrum_multivielbein} (their Eq. 3.22), as well as the set of algebraic constraints derived in terms of the non-symmetric $X_{\mu\nu}$ tensor\footnote{To first order in perturbation theory, this tensor is equivalent to $X^{(1)}_{\mu\nu}=\frac{\kappa}{2}\left[\delta g_{\mu\nu}-4\omega_{\mu\nu}\right]$ -- see appendix \ref{app:Volkov}.} in \cite{Volkov_arbitraryBG,Volkov_arbitraryBG_detailed} (Eq. 4.3 in \cite{Volkov_arbitraryBG_detailed}). If one is able to solve Eqs. \eqref{Linearised_Lorentz_constraints} for the Lorentz perturbations, then one is able to express the linearised field equations \eqref{GeneralLinearisedFieldEqs} in terms of only the metric perturbations, which is the ultimate goal.

It is worth pausing at this stage to ask the important question: what benefit is there to constructing the linearised field equations in this manner, as opposed to simply using the matrix technology that was previously derived in \cite{MG_arbitrary_BG,Syzygies,linear_spin2_gen_BG}, or using the non-symmetric tensor technology of \cite{Volkov_arbitraryBG,Volkov_arbitraryBG_detailed} (both of which we recap in appendix \ref{app:old perts}), since many of the expressions we have derived throughout this section appear quite complicated?

Regarding the matrix approach, to an extent one could argue, at least at linear level, that which formalism one uses is simply a matter of taste: as mentioned in the introduction, the potentially debilitating stage in the matrix calculation is that one needs to solve a challenging matrix equation in order to determine the linear perturbations to $S_{i\rightarrow j}$; in our formalism, this stage of the calculation is replaced by the requirement that one needs to solve the Lorentz constraints \eqref{Linearised_Lorentz_constraints}, which one might imagine could prove equally challenging around an arbitrary background. True as this might be, one should note that it is not necessary to solve for every $\omega^{(i)}$ individually -- just the differences $[\tensor{\omega}{^\mu_\nu}]_{i,j}$ suffice, as these are the combinations that appear in the dynamical equations for the metric perturbations. As we will see in the following section, around many physically relevant backgrounds, computing these differences is actually quite a simple task, certainly simpler than solving the analogous matrix equations. In fact, as we will demonstrate, on backgrounds with sufficient symmetry, it can be the case that the differences $[\tensor{\omega}{^\mu_\nu}]_{i,j}$ drop out of the field equations entirely, in which case simply knowing the form of the background $\bar{S}_{i\rightarrow j}$ matrices is all one needs to determine the perturbation structure -- our approach then becomes significantly simpler than the matrix approach.

Of course, the same could also be said of the approach from \cite{Volkov_arbitraryBG,Volkov_arbitraryBG_detailed} using the non-symmetric tensor $X_{\mu\nu}$, since there the difficulties associated with the matrix equations are also circumvented by working in terms of the vielbein perturbations. Indeed, in spirit, our approach and the one from \cite{Volkov_arbitraryBG,Volkov_arbitraryBG_detailed} are actually one and the same. However, there are two key practical differences between our formalisms that, at least in our opinion, make our approach the more useful of the two: firstly, we are able to explicitly parametrise the $X_{\mu\nu}$ tensor in terms of the metric perturbations and local Lorentz perturbations via Eq. \eqref{VielbeinPerturbations}, which makes eliminating the unphysical degrees of freedom (through the Lorentz constraints \eqref{Linearised_Lorentz_constraints}) more straightforward; secondly, we are able to easily compute the effective mass terms by simply contracting the relevant number of background $\bar{S}_{i\rightarrow j}$ matrices with generalised Kronecker deltas, where their approach instead involved solving a number of complicated algebraic equations to express the background $\bar{S}_{i\rightarrow j}$ matrices in terms of the background curvature quantities and $\beta_m^{(i,j)}$ parameters -- see appendix \ref{app:Volkov} for the details. Using their approach, it quickly becomes a challenge to compute these effective mass terms once additional dynamical metrics or extra dimensions are involved, whereas with our formalism this is a straightforward task, as we have demonstrated by doing all our calculations to this point in complete generality. Moreover, it is obvious in our formalism how to determine corrections beyond the linear order: one should simply include more terms from Eq. \eqref{VielbeinPerturbations} in the expansion of the vielbeins, which will lead to additional tensor structures emerging, taking a similar form to Eqs. \eqref{Atens}--\eqref{Ctens}, only containing more free indices (we will see this explicitly for the cubic terms in section \ref{sec:cubic}).

Finally, we would like to stress that the ghost freedom of the theory, which we know holds at the full nonlinear level, is clearly manifest in our formalism at the level of the perturbed action. Due to their structure in terms of the generalised deltas, the effective mass terms (involving the $\mathcal{B}$-tensors) will only contribute to the quadratic action a term that is \emph{linear} in $\tensor{\delta g}{^{(i)0}_0}$: any term involving $(\tensor{\delta g}{^{(i)0}_0})^2$ would require a delta with multiple 0 indices on one of its rows, which automatically vanishes by antisymmetry. Since the GR-like terms also only contribute a linear term in $\tensor{\delta g}{^{(i)0}_0}$ (as it is the linear analogue of the lapse function, which in standard GR acts as a Langrange multiplier), the full quadratic multi-gravity action also remains linear in $\tensor{\delta g}{^{(i)0}_0}$. Thus, $\tensor{\delta g}{^{(i)0}_0}$ enforces a primary constraint. On the other hand, while the GR-like terms are linear in $\tensor{\delta g}{^{(i)j}_0}$ (the linear analogue of the shift vector), the effective mass terms are quadratic in these variables, so they do not directly give rise to constraints. Instead, $\tensor{\delta g}{^{(i)j}_0}$ behave as auxiliary fields, which can be uniquely determined in terms of the genuinely dynamical variables and can hence be eliminated from the action\footnote{Really, one can only solve for $N-1$ of these, since they always appear in the effective mass terms via the difference combinations $[\tensor{\delta g}{^j_0}]_{i,j}$ (of which there are $N-1$ independent pairings, assuming no cycles -- see figure \ref{fig:interaction structure}). This is of course consistent with the fact that one of the spin-2 fields remains massless; the remaining $\dg{i}{j}{0}$ is a gauge mode of the surviving diagonal diffeomorphism invariance. The same is true of the $\dg{i}{0}{0}$: $N-1$ of them are Lagrange multipliers, and one is pure gauge.}; the resulting Hamiltonian after eliminating the auxiliary fields is second-class. Consequently, when one enforces the constancy of the primary constraint in time, it will inevitably lead to a further secondary, second-class constraint, in complete analogy with what happens in the usual FP story. This is how we see using our formalism that the perturbations of dRGT-type theories remain ghost-free, even around more generic backgrounds.

\section{Multi-gravity perturbations: examples}\label{sec:examples}

Let us now use the formalism we have developed to compute the structure of the multi-gravity perturbations around some commonly occurring example background spacetimes. We will begin by reproducing the perturbation structure around the so-called \emph{proportional solutions}, the simplest vacuum solutions of multi-gravity, before moving on to look at the perturbation structure around cosmological and black hole backgrounds.

\subsection{Proportional backgrounds}\label{sec:prop}

Let us work in vacuum, with $I_M=0$ for simplicity. As the name suggests, the proportional solutions are those where all of the various metrics are proportional to one another \cite{prop_bg_multigrav,consistent_spin2}:
\begin{equation}\label{propmetrics}
    \gi{i} = a_i^2 \bar{g}_{\mu\nu} \; .
\end{equation}
where $\bar{g}_{\mu\nu}$ solves the vacuum field equations of GR with cosmological constant $\bar{\Lambda}$, and $a_i$ are a set of conformal factors that the Bianchi constraint forces to be constant \cite{consistent_spin2}.

With this ansatz, the relevant background curvature tensors are:
\begin{align}
    \tensor{\bar{R}}{^{(i)\mu}_\nu} &= \frac{\bar{R}^\mu_{\;\nu}}{a_i^2} \; , \;\;\; \bar{R}^{(i)}=\frac{\bar{R}}{a_i^2} \; , \;\;\; \tensor{\bar{G}}{^{(i)\mu}_\nu} = \frac{\bar{G}^\mu_{\;\nu}}{a_i^2} \; \nonumber
    \\
    \bar{R}_{\mu\nu} &= \frac{2\bar{\Lambda}}{D-2}\bar{g}_{\mu\nu} \; , \;\;\; \bar{R} = \frac{2D\bar{\Lambda}}{D-2} \; , \;\;\; \bar{G}_{\mu\nu} = -\bar{\Lambda}\bar{g}_{\mu\nu} \; ,\label{BGcurvature}
\end{align}
and the building-block matrices are simply $(\Sij{i}{j})^\mu_{\;\nu}=(a_j/a_i)\delta^\mu_\nu$, leading to the following form for the $W$-tensor components upon substitution into Eq. \eqref{W_metric}:
\begin{align}\label{W_comps_prop}
    \Wi{i} = \delta^\mu_\nu&\Bigg[\sum_j\sum_{m=0}^{D} \beta_m^{(i,j)} \binom{D-1}{m}\left(\frac{a_j}{a_i}\right)^m \nonumber
    \\
    &+ \sum_k\sum_{m=0}^{D} \beta_{D-m}^{(k,i)} \binom{D-1}{m} \left(\frac{a_k}{a_i}\right)^m \Bigg]\; .
\end{align}
Consequently, the multi-gravity vacuum equations $M_i^{D-2}\Gi{i}+\Wi{i}=0$ become the following $N$ algebraic, nonlinear simultaneous equations:
\begin{equation}\label{propsols}
\begin{split}
    \frac{\bar{\Lambda}M_i^{D-2}}{a_i^2} = &\sum_j \sum_{m=0}^{D} \beta_m^{(i,j)} \binom{D-1}{m}\left(\frac{a_j}{a_i}\right)^m 
    \\
    &+ \sum_k \sum_{m=0}^{D} \beta_{D-m}^{(k,i)} \binom{D-1}{m} \left(\frac{a_k}{a_i}\right)^m \; ,
\end{split}
\end{equation}
which, after fixing one of the $a_i$ via coordinate rescaling, may be solved for $\bar{\Lambda}$ and the remaining $N-1$ conformal factors, the physical solutions being those with real $\bar{\Lambda}$ and $a_i$. In this way, multi-gravity naturally admits de Sitter (dS), anti-de Sitter (AdS) and Minkowski vacua, where the interactions between metrics manifest themselves as an effective cosmological constant.

Turning to the perturbations, let us first consider the linearised Lorentz constraints \eqref{Linearised_Lorentz_constraints}. The fact that $\bar{S}_{i\rightarrow j}$ is proportional to the identity matrix hugely simplifies things here, as it implies that $S$ and $\delta g$ commute, so the metric perturbations drop out of Eq. \eqref{Linearised_Lorentz_constraints}\footnote{In fact, this can \emph{only} happen around proportional backgrounds, as only when $S\propto\mathbbm{1}$ does $S$ commute with a generic $\delta g$.}. Thus, the Lorentz constraints are trivially solved by:
\begin{equation}
    \kappa_i\om{i}{\mu}{\nu} = \kappa_j\om{j}{\mu}{\nu} \implies [\tensor{\omega}{^\mu_\nu}]_{i,j} = 0 \; ,
\end{equation}
for all pairs $i,j$. Consequently, the local Lorentz fields drop out of the field equations \eqref{GeneralLinearisedFieldEqs} for the metric perturbations entirely -- this solution for the $\omega^{(i)}$'s corresponds to performing an overall Lorentz transformation on all vielbeins, which is a symmetry of the theory.

Moving on to the metric equations \eqref{GeneralLinearisedFieldEqs}, using the background curvature tensors from Eqs. \eqref{BGcurvature}, the two kinetic terms become:
\begin{equation}\label{PropKinetic}
    \begin{split}
        &\tensor{\bar{\mathcal{E}}}{^{(i)\mu}_\nu^\alpha_\beta}\tensor{\delta g}{^{(i)\beta}_\alpha} + \tensor{\mathcal{R}}{^{(i)\mu}_\nu^\alpha_\beta}\tensor{\delta g}{^{(i)\beta}_\alpha} 
        \\
        &\;= \frac{\tensor{\bar{\mathcal{E}}}{^{\mu}_\nu^\alpha_\beta}\tensor{\delta g}{^{(i)\beta}_\alpha}}{a_i^2} - \frac{\bar{\Lambda}}{a_i^2(D-2)}\left(D\tensor{\delta g}{^{(i)\mu}_\nu}-\delta^\mu_\nu\delta g^{(i)}\right)
        \\
        &\qquad+ \frac{\bar{\Lambda}}{a_i^2}\left(2\tensor{\delta g}{^{(i)\mu}_\nu}-\frac12\delta^\mu_\nu\delta g^{(i)}\right) \; .
    \end{split}
\end{equation}
Furthermore, from the background field equations \eqref{propsols} we have $\tensor{\bar{W}}{^{(i)\mu}_\nu}=\delta^\mu_\nu M_i^{D-2}\bar{\Lambda}/a_i^2$, and hence:
\begin{equation}\label{PropW}
\begin{split}
    &\frac{\kappa_i^2}{2}\left[\tensor{\bar{W}}{^{(i)\mu}_\lambda}\tensor{\delta g}{^{(i)\lambda}_\nu} + \tensor{\bar{W}}{^{(i)\lambda}_\nu}\tensor{\delta g}{^{(i)\mu}_\lambda}-\tensor{\bar{W}}{^{(i)\mu}_\nu}\delta g^{(i)}\right] 
    \\
    &= \frac{\bar{\Lambda}}{a_i^2}\left[\tensor{\delta g}{^{(i)\mu}_\nu}-\frac12\delta^\mu_\nu\delta g^{(i)}\right] \; .
\end{split}
\end{equation}
Regarding the mass terms, the fact that $\Sij{i}{j}\propto\mathbbm{1}$ again simplifies things significantly, as all contractions with the generalised deltas in the $\mathcal{B}$-tensors may be evaluated using Eq. \eqref{ProductDeltas} to yield:
\begin{align}
    [\mathcal{B}_{i,j}^{(+)}]^{\mu\;\;\alpha}_{\;\;\nu\;\;\beta} &= \sigma_{i,j}^{(+)}\delta^{\mu\alpha}_{\nu\beta} \; ,\label{PropB1}
    \\
    [\mathcal{B}_{i,k}^{(-)}]^{\mu\;\;\alpha}_{\;\;\nu\;\;\beta}  &= \sigma_{i,k}^{(-)}\delta^{\mu\alpha}_{\nu\beta} \; ,\label{PropB2}
\end{align}
where we have defined the following combinations of interaction coefficients and conformal factors:
\begin{align}
    \sigma^{(+)}_{i,j} &=  \sum_{m=0}^D \beta_m^{(i,j)}\binom{D-2}{m-1}\left(\frac{a_j}{a_i}\right)^{m-1} \; ,\label{sig_p}
    \\
    \sigma^{(-)}_{i,k} &= \sum_{m=0}^D \beta_{D-m}^{(k,i)}\binom{D-2}{m-1}\left(\frac{a_k}{a_i}\right)^{m-1} \; .\label{sig_m}
\end{align}
These two parameters appear ubiquitously in multi-gravity -- we will see them appearing in cosmological and black hole perturbations later on as well. The plus and minus variants associated to any given pair of metrics are related to one another by:
\begin{equation}\label{Sig_pm}
     \sigma^{(-)}_{j,i} = \left(\frac{a_{i}}{a_{j}}\right)^{D-2} \sigma^{(+)}_{i,j} \; ,
\end{equation}
informing us that the $\sigma_{i,j}^{(\pm)}$ parameters live on the interaction links coupling $\gi{i}$ and $\gi{j}$.

Substituting Eqs. \eqref{PropKinetic}, \eqref{PropW}, \eqref{PropB1} and \eqref{PropB2} into Eqs. \eqref{GeneralLinearisedFieldEqs}, one finds that the linearised field equations around a proportional background are:
\begin{equation}
\begin{split}
    &\frac{\tensor{\bar{\mathcal{E}}}{^{\mu}_{\nu}^\alpha_\beta}\delta g^{(i)\beta}_{\;\;\;\;\;\alpha}}{a_i^2} - \frac{2\bar{\Lambda}}{a_i^2(D-2)}\left(\tensor{\delta g}{^{(i)\mu}_{\nu}}-\frac12\delta^\mu_\nu\delta g^{(i)}\right)
    \\
    &+ \frac12 \Bigg\{\sum_j \kappa_i\frac{a_j}{a_i}\sigma_{i,j}^{(+)}\bigg(\delta^\mu_\nu[\delta g]_{i,j}-[\tensor{\delta g}{^\mu_\nu}]_{i,j}\bigg)
    \\
    &\qquad+ \sum_k \kappa_i\frac{a_k}{a_i}\sigma_{i,k}^{(-)}\bigg(\delta^\mu_\nu[\delta g]_{i,k}-[\tensor{\delta g}{^\mu_\nu}]_{i,k}\bigg)\Bigg\} = 0 \; .
\end{split}
\end{equation}
We can rewrite this so that all tensorial quantities behave as if they lived in the \emph{common} background of $\bar{g}_{\mu\nu}$, by defining:
\begin{equation}
    \tensor{\delta\tilde{g}}{^{(i)\mu}_\nu} \equiv a_i^2\tensor{\delta g}{^{(i)\mu}_\nu} = \bar{g}^{\mu\lambda}\delta g^{(i)}_{\lambda\nu} \; ,
\end{equation}
which has its indices manipulated with $\bar{g}_{\mu\nu}$, rather than $g^{(i)}_{\mu\nu}$. Rewriting the above set of field equations in terms of the tilded metric perturbations, then lowering an index using $\bar{g}_{\mu\nu}$, one arrives at:
\begin{equation}\label{LinEFEsNonCanonical}
\begin{split}
    &\tensor{\bar{\mathcal{E}}}{_{\mu\nu}^\alpha_\beta}\delta \tilde{g}^{(i)\beta}_{\;\;\;\;\;\alpha}- \frac{2\bar{\Lambda}}{D-2}\left(\delta \tilde{g}^{(i)}_{\mu\nu}-\frac12\bar{g}_{\mu\nu}\delta \tilde{g}^{(i)}\right)
    \\
    &+ \frac{a_i^2}{2} \Bigg\{\sum_j \kappa_i \sigma_{i,j}^{(+)}\bigg[\kappa_i\frac{a_j}{a_i}\left(\delta\tilde{g}^{(i)}_{\mu\nu}-\bar{g}_{\mu\nu}\delta \tilde{g}^{(i)}\right) 
    \\
    &\qquad\qquad\qquad\qquad- \kappa_j\frac{a_i}{a_j}\left(\delta\tilde{g}^{(j)}_{\mu\nu}-\bar{g}_{\mu\nu}\delta \tilde{g}^{(j)}\right)\bigg]
    \\
    &\qquad+ \sum_k \kappa_i\sigma_{i,k}^{(-)}\bigg[\kappa_i\frac{a_k}{a_i}\left(\delta\tilde{g}^{(i)}_{\mu\nu}-\bar{g}_{\mu\nu}\delta \tilde{g}^{(i)}\right) 
    \\
    &\qquad\qquad\qquad\qquad- \kappa_k\frac{a_i}{a_k}\left(\delta\tilde{g}^{(k)}_{\mu\nu}-\bar{g}_{\mu\nu}\delta \tilde{g}^{(k)}\right)\bigg]\Bigg\} = 0\; .
\end{split}
\end{equation}
One can check that these equations agree with the linearised multi-metric field equations given in \cite{prop_bg_multigrav}, for the two classes of theory graph they consider (`star-type' and `chain-type' interactions); our Eqs. \eqref{LinEFEsNonCanonical} constitute the appropriate generalisation to arbitrary interaction structures with any number of positively and negatively oriented interactions per metric.

A final step we can take, given that we have already rewritten everything with respect to the common background metric $\bar{g}_{\mu\nu}$, is to ensure that the perturbations are all canonically normalised with respect to $\bar{g}_{\mu\nu}$. This requires that we take:
\begin{equation}\label{CanonicalNormalisation}
    \delta \tilde{g}^{(i)}_{\mu\nu}=a_i h^{(i)}_{\mu\nu} \; ,
\end{equation}
With this choice, the linearised field equations \eqref{LinEFEsNonCanonical} become:
\begin{equation}\label{LinEFEsCanonical}
\begin{split}
    \tensor{\bar{\mathcal{E}}}{_{\mu\nu}^\alpha_\beta} h^{(i)\beta}_{\;\;\;\;\;\alpha} &- \frac{2\bar{\Lambda}}{D-2}\left(h^{(i)}_{\mu\nu}-\frac12\bar{g}_{\mu\nu}h^{(i)}\right) 
    \\
    &+ \frac{\mathcal{M}_{ij}^2}{2}\left( h^{(j)}_{\mu\nu}-\bar{g}_{\mu\nu} h^{(j)}\right) = 0 \; ,
\end{split}
\end{equation}
which is precisely in the multi-FP form alluded to by Eq. \eqref{MultiFP}, where the mass matrix has components:
\begin{align}
    \mathcal{M}_{ii}^2 &= \kappa_i^2 a_i^2 \left(\sum_j \frac{a_{j}}{a_i}\sigma_{i,j}^{(+)} + \sum_k\frac{a_{k}}{a_i}\sigma_{i,k}^{(-)}\right) \label{Mii} \; ,
    \\
    \mathcal{M}_{ji}^2&=\left(\frac{a_{j}}{a_i}\right)^{4-D}\mathcal{M}_{ij}^2 = -\kappa_i\kappa_j a_i^2 \sigma_{i,j}^{(+)}\label{Mi1} \; ,
\end{align}
and all tensorial quantities behave as if they lived in the common background of $\bar{g}_{\mu\nu}$. It is a simple check to show that there always exists a single massless eigenvector (i.e. with eigenvalue $m_0^2=0$) scaling as $O^{j0}\propto a_j/\kappa_j$, irrespective of any choices for $\beta_m^{(i,j)}$, corresponding to the surviving diagonal copy of diffeomorphism invariance in the theory. This is in accordance with the no-go theorem mentioned in the introduction forbidding theories of multiple interacting massless spin-2 fields \cite{no_interacting_massless_gravitons}.

\subsection{Cosmological backgrounds}

Now let us turn to a background in which the perturbation structure is somewhat more involved, namely, that of an expanding FLRW universe, where the most general homogeneous and isotropic ansatz one can make for the metrics at background level (in $D=4$ dimensions) is:
\begin{equation}\label{FLRWMultiGrav}
    \gi{i}\dd x^\mu \dd x^\nu = -c_i^2(t) \dd t^2 + a_i^2(t)\left[\frac{\dd r^2}{1-kr^2} + r^2\dd\Omega_2^2\right] \; ,
\end{equation}
where $c_i(t)$ are the \emph{lapses}, $a_i(t)$ are the \emph{scale factors}, $\dd\Omega_2^2$ is the line element for a 2-sphere, and it can be shown that the Bianchi constraints \eqref{Bianchi constraint} force the spatial curvature $k$ to be the same for all metrics \cite{k=0}. One of the lapses, say, on the distinguished metric $\gi{i_*}$, can be fixed to $c_{i_*}(t)=1$ throughout all time by choosing a preferred time coordinate; ordinarily one would choose $\gi{i_*}$ to be the metric to which matter couples, so that $t$ then corresponds to the cosmic time.

The background cosmology of multi-gravity, with the ansatz \eqref{FLRWMultiGrav} for the metrics, is discussed in detail in \cite{KWThesis}. Essentially, solutions split into three branches depending on how the Bianchi constraints are satisfied, of which two are unstable to nonlinear ghosts and growing tensor modes \cite{nonlinear_instability_FRW,nonlinear_cosmo_stability,general_mass}, but the third (which requires $c_i=\dot{a}_i/\dot{a}_{i_*}$) is stable and has particularly interesting (and potentially observationally relevant \cite{analytical_constraints_bigravity,observational_constraints_bigravity,combining_cosmo_local_bounds_bigravity,HubbleTensionBigravity,Bigravity_DESI,Bigravity_Hubble_DESI}) phenomenology e.g. there is a dynamical effective dark energy component arising due to the $W$-tensors that has a phantom equation of state in the distant past but which later relaxes to an effective cosmological constant once all external matter has diluted away.

For our purposes, all we wish to do is to use our perturbation formalism to compute the effective mass terms around this FLRW background. To this end, immediately from Eq. \eqref{FLRWMultiGrav} one sees that the building-block matrices $\Sij{i}{j}$ take the simple form:
\begin{equation}\label{S_Cosmo}
     S_{i\rightarrow j} = \text{diag}\left(\frac{c_j}{c_i}, \frac{a_j}{a_i}, \frac{a_j}{a_i}, \frac{a_j}{a_i}\right) \; .
\end{equation}
Substituting into Eq. \eqref{Linearised_Lorentz_constraints} leads to the following Lorentz constraints:
\begin{align}
    [\tensor{\omega}{^m_n}]_{i,j} &= 0 \; ,\label{CosmoLorentz1}
    \\
    4\left(\frac{a_j}{a_i}+\frac{c_j}{c_i}\right)[\tensor{\omega}{^0_m}]_{i,j} &= -\left(\frac{a_j}{a_i}-\frac{c_j}{c_i}\right)[\tensor{\delta g}{^0_m}]_{i,j} \; ,\label{CosmoLorentz}
\end{align}
which are naturally split in a 3+1 manner owing to the presence of the lapses. The spatial components, Eqs. \eqref{CosmoLorentz1}, are solved trivially and hence drop out of the metric field equations, thanks to the proportionality of the spatial metrics. However, the components mixing space and time indices, Eqs. \eqref{CosmoLorentz}, are non-trivially solved, and will hence contribute to the metric equations for $\dg{i}{0}{m}$.

Regarding said metric equations, by substituting Eq. \eqref{S_Cosmo} into Eqs. \eqref{Btens} and \eqref{Ctens}, one finds the following structure for the effective spin-2 mass terms:
\begin{align}
    &[\mathcal{B}_{i,j}^{(+)}]^{0\;\;\alpha}_{\;\;0\;\;\beta}(\bar{S}_{i\rightarrow j})^\beta_{\;\lambda}\tensor{\chi}{^{\lambda}_\alpha} = \frac{a_j}{a_i}\sigma_{i,j}^{(+)}(t)\left(\delta^0_0 {\chi}-\tensor{\chi}{^{0}_0}\right) \; ,\label{CosmoPertsMultiGravity1}
    \\
    &[\mathcal{B}_{i,j}^{(+)}]^{m\;\;\alpha}_{\;\;n\;\;\beta}(\bar{S}_{i\rightarrow j})^\beta_{\;\lambda}\tensor{\chi}{^{\lambda}_\alpha} = \frac{c_j}{c_i}\sigma_{i,j}^{(+)}(t)\tensor{\chi}{^{0}_0}\delta^m_n\nonumber
        \\ &\qquad\qquad\qquad\qquad+\frac{a_j}{a_i}\varsigma_{i,j}^{(+)}(t)\left(\delta^m_n {\chi}^{k}_{\;k}-\tensor{\chi}{^{m}_n}\right) \; ,
    \\
    &[\mathcal{B}_{i,j}^{(+)}]^{0\;\;\alpha}_{\;\;m\;\;\beta}(\bar{S}_{i\rightarrow j})^\beta_{\;\lambda}\tensor{\chi}{^{\lambda}_\alpha} = -\frac{c_j}{c_i}\sigma_{i,j}^{(+)}(t)\tensor{\chi}{^{0}_m} \; ,
    \\
    &[\mathcal{B}_{i,j}^{(+)}]^{m\;\;\alpha}_{\;\;0\;\;\beta}(\bar{S}_{i\rightarrow j})^\beta_{\;\lambda}\tensor{\chi}{^{\lambda}_\alpha} = -\frac{a_j}{a_i}\sigma_{i,j}^{(+)}(t)\tensor{\chi}{^{m}_0} \; ,
\end{align}
for the positively oriented interactions, and:
\begin{align}
    &[\mathcal{B}_{i,k}^{(-)}]^{0\;\;\alpha}_{\;\;0\;\;\beta}(\bar{S}_{i\rightarrow k})^\beta_{\;\lambda}\tensor{\chi}{^{\lambda}_\alpha} = \frac{a_k}{a_i}\sigma_{i,k}^{(-)}(t)\left(\delta^0_0 {\chi}-\tensor{\chi}{^{0}_0}\right) \; ,
    \\
    &[\mathcal{B}_{i,k}^{(-)}]^{m\;\;\alpha}_{\;\;n\;\;\beta}(\bar{S}_{i\rightarrow k})^\beta_{\;\lambda}\tensor{\chi}{^{\lambda}_\alpha} = 
        \frac{c_k}{c_i}\sigma_{i,k}^{(-)}(t)\tensor{\chi}{^{0}_0}\delta^m_n\nonumber
        \\ 
    &\qquad\qquad\qquad\qquad+\frac{a_k}{a_i}\varsigma_{i,k}^{(-)}(t)\left(\delta^m_n {\chi}^{k}_{\;k}-\tensor{\chi}{^{m}_n}\right) \; ,
    \\
    &[\mathcal{B}_{i,k}^{(-)}]^{0\;\;\alpha}_{\;\;m\;\;\beta}(\bar{S}_{i\rightarrow k})^\beta_{\;\lambda}\tensor{\chi}{^{\lambda}_\alpha} = -\frac{c_k}{c_i}\sigma_{i,k}^{(-)}(t)\tensor{\chi}{^{0}_m} \; ,
    \\
    &[\mathcal{B}_{i,k}^{(-)}]^{m\;\;\alpha}_{\;\;0\;\;\beta}(\bar{S}_{i\rightarrow k})^\beta_{\;\lambda}\tensor{\chi}{^{\lambda}_\alpha} = -\frac{a_k}{a_i}\sigma_{i,k}^{(-)}(t)\tensor{\chi}{^{m}_0} \; ,\label{CosmoPertsMultiGravity2}
\end{align}
for the negatively oriented interactions, where we have defined the following time-dependent parameters:
\begin{align}
    \sigma_{i,j}^{(+)}(t) &= \beta_1^{(i,j)} + 2\beta_2^{(i,j)}\frac{a_j}{a_i} + \beta_3^{(i,j)}\left(\frac{a_j}{a_i}\right)^2\label{sigp_t} \; ,
    \\
    \sigma_{i,k}^{(-)}(t) &= \beta_3^{(k,i)} + 2\beta_2^{(k,i)}\frac{a_k}{a_i} + \beta_1^{(k,i)}\left(\frac{a_k}{a_i}\right)^2\label{sigm_t} \; ,
    \\
    \varsigma_{i,j}^{(+)}(t) &= \beta_1^{(i,j)} + \beta_2^{(i,j)}\left(\frac{c_j}{c_i}+\frac{a_j}{a_i}\right)+\beta_3^{(i,j)}\frac{c_j}{c_i}\frac{a_j}{a_i} \; ,\label{varsigp}
    \\
    \varsigma_{i,k}^{(-)}(t) &= \beta_3^{(k,i)} + \beta_2^{(k,i)}\left(\frac{c_k}{c_i}+\frac{a_k}{a_i}\right)+\beta_1^{(k,i)}\frac{c_k}{c_i}\frac{a_k}{a_i} \; ,\label{varsigm}
\end{align}
which are related by:
\begin{align}
    \sigma_{i,j}^{(+)}(t) &= \left(\frac{a_j}{a_i}\right)^2 \sigma_{j,i}^{(-)}(t) \; ,
    \\
    \varsigma_{i,j}^{(+)}(t) &= \frac{c_j}{c_i}\frac{a_j}{a_i}\varsigma_{j,i}^{(-)}(t) \; .
\end{align}
We note that the $\sigma_{i,j}^{(\pm)}(t)$ parameters are simply time-dependent analogues of the constant $\sigma_{i,j}^{(\pm)}$ parameters we introduced in Eqs. \eqref{sig_p} and \eqref{sig_m} to encode the spin-2 mass matrix around a proportional background, where the constant conformal factors have been replaced by the time-dependent scale factors. The $\varsigma_{i,j}^{(\pm)}$ parameters have no such analogue in general, but note that if $a_i=c_i$ (which happens at late times after all external matter has diluted away \cite{analytical_constraints_bigravity}) then the two metrics are actually proportional; in this case, $\sigma_{i,j}^{(\pm)}(t)$ and $\varsigma_{i,j}^{(\pm)}(t)$ coincide, and the perturbation structure of Eqs. \eqref{CosmoPertsMultiGravity1}--\eqref{CosmoPertsMultiGravity2} reduces down to that of Eqs. \eqref{PropB1} and \eqref{PropB2} around a proportional solution, as it should do (of course, the non-trivial Lorentz constraints \eqref{CosmoLorentz} also become trivial in this scenario, as the expression on the right hand side vanishes).

One can check that the effective mass terms in Eqs. \eqref{CosmoPertsMultiGravity1}--\eqref{CosmoPertsMultiGravity2} agree with those appearing in the linearised cosmological field equations of e.g. \cite{structure_growth}, upon splitting up the components of $\delta g^{(i)\mu}_{\;\;\;\;\;\nu}$ into their scalar, vector and tensor parts (the remaining terms in the linearised field equations are all GR-like terms coming from the kinetic sector, whose computation is well-understood -- see any classic cosmology textbook for details e.g. \cite{WeinbergCosmology} -- so we omit them here). In particular, in the $\dg{i}{0}{m}$ equations, for which the Lorentz constraints are non-trivial, the effective mass term (written explicitly for a positively oriented interaction) is:
\begin{align}
    &[\mathcal{B}_{i,j}^{(+)}]^{0\;\;\alpha}_{\;\;m\;\;\beta}(\bar{S}_{i\rightarrow j})^\beta_{\;\lambda} \left\{[\tensor{\delta g}{^{\lambda}_\alpha}]_{i,j}-4[\tensor{\omega}{^\lambda_\alpha}]_{i,j}\right\} \nonumber
    \\
    &\qquad\qquad\qquad\qquad= -2\frac{\frac{a_j}{a_i}\frac{c_j}{c_i}}{\frac{a_j}{a_i}+\frac{c_j}{c_i}}\sigma_{i,j}^{(+)}(t)[\tensor{\delta g}{^0_m}]_{i,j} \; ,
\end{align}
which one can check agrees with Eq. (A.4) in \cite{structure_growth} upon parametrising the metric perturbations accordingly. 

The increased generality of our approach provides a number of benefits over previous matrix approaches, for example: a nice application of Eqs. \eqref{CosmoPertsMultiGravity1}--\eqref{CosmoPertsMultiGravity2} is that one can derive the Higuchi bound \cite{OG_Higuchi,OG_Higuchi_2,General_Higuchi_bound,Higuchi_bound_massive_grav,higuchi_and_gradient} on cosmological spacetimes in a very neat and clean manner, as well as see very clearly why certain branches of cosmological solutions in multi-gravity are unstable -- see \cite{KWThesis} for details.

\subsection{Black hole backgrounds}

The final example we would like to give concerns the structure of the effective mass terms around black hole backgrounds. Such calculations are important when one wishes to compute e.g. the quasinormal spectrum of a black hole in massive gravity, with a view to comparison against gravitational wave measurements of black hole binaries. To this end, a complete catalogue of the known black hole solutions in generic multi-metric theories was constructed in \cite{BHs_multigrav,BH2}; all such solutions that may be written down analytically can be cast into Kerr-Schild form:
\begin{equation}\label{KdSMultiGrav}
   \gi{i} = a_i^2\left[g^{(\Lambda)}_{\mu\nu} + 2\phi_i l_\mu l_\nu\right] \; ,
\end{equation}
where $a_i$ are again constant conformal factors, $g^{(\Lambda)}_{\mu\nu}$ is the metric of AdS, Minkowski or dS space ($\Lambda<0$, $\Lambda=0$, $\Lambda>0$, respectively) in some coordinate system, $\phi_i$ are scalar functions containing the Schwarzschild radii for each metric (which are in principle independent at this stage) and $l$ is a vector tangent to a null-geodesic congruence on $g^{(\Lambda)}_{\mu\nu}$ -- see \cite{BHs_multigrav} for the explicit expressions. The ansatz \eqref{KdSMultiGrav} is able to encompass black holes of arbitrary dimension that can rotate in multiple planes; in $D=4$, or without any rotation, it can also account for electric charge. 

As in the cosmological case, there are three distinct branches of solutions (this time depending on whether the metrics are simultaneously diagonalisable) and as before two of these are expected to be pathological: they are analogous to the two pathological branches of cosmological solutions in multiple ways, and we in fact believe the underlying source of the pathology to be the same in both cases -- see \cite{KWThesis,BH2} for a discussion. However, the pathology has not yet been confirmed explicitly, and it is likely that one will need to go to nonlinear order in perturbation theory to see it.

In any case, with the multi-gravity metrics given in the Kerr-Schild form \eqref{KdSMultiGrav}, the fact that $l$ is null ($l_\mu l^\mu=0$) makes it surprisingly simple to calculate the form of the $S_{i\rightarrow j}$ matrices, as there is an early truncation in the expansion of the matrix square root \cite{Rotating_AdS_bigravity}:
\begin{equation}\label{BH_Sij}
    (S_{i\rightarrow j})^\mu_{\;\nu} = \frac{a_j}{a_i}\left[\delta^\mu_\nu - (\phi_i-\phi_j)l^\mu l_\nu\right] \; .
\end{equation}
Substituting into Eqs. \eqref{Btens} and \eqref{Ctens} and again utilising the null character of $l$, one finds the following perturbation structure at linear order:
\begin{align}
    &[\mathcal{B}_{i,j}^{(+)}]^{\mu\;\;\alpha}_{\;\;\nu\;\;\beta} = \sigma_{i,j}^{(+)}\delta^{\mu\alpha}_{\nu\beta} \nonumber
    \\
    &+ 2\frac{a_j}{a_i}\left(\phi_i-\phi_j\right)\left(\beta_2^{(i,j)}+\frac{a_j}{a_i}\beta_3^{(i,j)}\right)\left(\delta^{[\mu}_\nu l^{\alpha]} l_\beta + \delta^{[\alpha}_\beta l^{\mu]} l_\nu \right)\; ,\label{BHperts1}
    \\
    &[\mathcal{B}_{i,k}^{(-)}]^{\mu\;\;\alpha}_{\;\;\nu\;\;\beta} = \sigma_{i,k}^{(-)}\delta^{\mu\alpha}_{\nu\beta} \nonumber
    \\
    &+ 2\frac{a_k}{a_i}\left(\phi_i-\phi_k\right)\left(\beta_2^{(k,i)}+\frac{a_k}{a_i}\beta_1^{(k,i)}\right)\left(\delta^{[\mu}_\nu l^{\alpha]} l_\beta + \delta^{[\alpha}_\beta l^{\mu]} l_\nu \right)\; .\label{BHperts2}
\end{align}
as well as the following set of Lorentz constraints:
\begin{align}\label{BH_lorentz}
    4[\omega_{\mu\nu}]_{i,j} = (\phi_i-\phi_j)\left\{[\delta g_{\lambda[\mu}]_{i,j}l^\lambda l_{\nu]} - 4[\omega_{\lambda[\mu}]_{i,j}l^\lambda l_{\nu]}\right\} \; .
\end{align}

If all the $\phi_i$ are equal, then clearly all the metrics in Eq. \eqref{KdSMultiGrav} are proportional, and Eqs. \eqref{BHperts1} and \eqref{BHperts2} recover the corresponding Eqs. \eqref{PropB1} and \eqref{PropB2} for the $\mathcal{B}$-tensors around a proportional solution, as they should. Likewise, the linearised Lorentz constraints \eqref{BH_lorentz} reduce to $[\omega_{\mu\nu}]_{i,j}=0$, as befitting a proportional solution. When $\phi_i\neq\phi_j$, the perturbation structure and linearised Lorentz constraints are much more involved: in fact, the linearised field equations around non-proportional black hole solutions in multi-gravity have only previously been computed for a background that is 4-dimensional and non-rotating \cite{GL_unified,Stability_nonbidiag,BH2}. We will show in appendix \ref{app:BH} how to recover those results from our Eqs. \eqref{BHperts1} and \eqref{BHperts2}, but one should note that our expressions are far more general: they hold in arbitrary dimension, and constitute the first time the effective mass term around a non-proportional \emph{rotating} black hole background in multi-gravity has been computed, demonstrating the power of our formalism for determining the perturbations. 

\vspace{-0.8em}
\section{Cubic potential}\label{sec:cubic}

Finally, we are going to determine the multi-gravity potential to cubic order around an arbitrary background; previously, this calculation has only been performed around a proportional background (see section \ref{sec:examples}) in $D=4$ \cite{heavy_spin2_DM}, so here we again chart new territory. 

The cubic order variation of the potential, considering all terms from the expansion \eqref{1formperts} that are third-order in the metric perturbations, has 3 contributions: \begin{align}
    I_V^{(3)} = &-\sum_{ijkl_4\hdots l_{D}} \int\kappa_i\kappa_j\kappa_k \binom{D}{3}T_{ijkl_4\hdots l_{D}} \varepsilon_{abcd_4\hdots d_{D}} \nonumber
    \\
    &\times\delta e^{(i)a}_{(1)}\wedge \delta e^{(j)b}_{(1)}\wedge \delta e^{(k)c}_{(1)}\wedge \bar{e}^{(l_4)d_4}\wedge\hdots\wedge\bar{e}^{(l_{D})d_{D}} \nonumber
    \\
    &-\sum_{ijl_3\hdots l_{D}}\int\kappa_i^2\kappa_j\binom{D}{2} T_{ijl_3\hdots l_D}\varepsilon_{abd_3\hdots d_D}\nonumber
    \\
    &\qquad\times\delta e^{(i)a}_{(2)} \wedge \delta e^{(j)b}_{(1)}\wedge \bar{e}^{(l_3)d_3}\wedge\hdots\wedge\bar{e}^{(l_D)d_D}\nonumber
    \\
    &-\sum_{il_2\hdots l_D}\int \kappa_i^3 \binom{D}{1} T_{il_2\hdots l_D} \varepsilon_{ad_2\hdots d_D}\nonumber
    \\
    &\qquad\times \delta e^{(i)a}_{(3)} \wedge \bar{e}^{(l_2)d_2}\wedge\hdots\wedge\bar{e}^{(l_D)d_D}\; ,
\end{align}
which can be expanded out using Eqs. \eqref{1storder}--\eqref{3rdorder} to yield:
\begin{align}
    &I_V^{(3)} = -\sum_{ijkl} \int\dd[D]x\sqrt{-\det\bar{g}_{(i)}} \nonumber \;\; \times
    \\
    &\Bigg[\frac{\kappa_i\kappa_j\kappa_k}{8}\binom{D}{3}T_{ijkl_4\hdots l_D} \bar{e}^{(i)\mu\nu\rho\lambda_4\hdots\lambda_D}_{\;\;\;\;abcd_4\hdots d_D} \nonumber
    \\
    &\qquad\times (\delta g^{(i)}-4\omega^{(i)})^\alpha_{\;\mu}(\delta g^{(j)}-4\omega^{(j)})^\beta_{\;\nu}(\delta g^{(k)}-4\omega^{(k)})^\gamma_{\;\rho}\nonumber
    \\
    &\qquad\times\bar{e}^{(i)a}_\alpha \bar{e}^{(j)b}_\beta \bar{e}^{(k)c}_\gamma \bar{e}^{(l_4)d_4}_{\lambda_4}\hdots\bar{e}^{(l_D)d_D}_{\lambda_D}\nonumber
    \\
    &-\frac{\kappa_i^2\kappa_j}{16}\binom{D}{2} T_{ijl_3\hdots l_D} \bar{e}^{(i)\mu\nu\lambda_3\hdots\lambda_D}_{\;\;\;\;abd_3\hdots d_D}\nonumber
    \\
    &\qquad\times(\delta g^{(i)2}+8\delta g^{(i)}\omega^{(i)}-16\omega^{(i)2})^\alpha_{\;\mu}(\delta g^{(j)}-4\omega^{(j)})^\beta_{\;\nu}\nonumber
    \\
    &\qquad\times\bar{e}^{(i)a}_\alpha \bar{e}^{(j)b}_\beta\bar{e}^{(l_3)d_3}_{\lambda_3}\hdots\bar{e}^{(l_D)d_D}_{\lambda_D}\nonumber
    \\
    &-\frac{\kappa_i^3}{16}\binom{D}{1} T_{il_2\hdots l_D} \bar{e}^{(i)\mu\lambda_2\hdots\lambda_D}_{\;\;\;\;ad_2\hdots d_D}\nonumber
    \\
    &\qquad\times (\delta g^{(i)3}-4\delta g^{(i)2}\omega^{(i)}+16\delta g^{(i)}\omega^{(i)2}+32\omega^{(i)3})^\alpha_{\;\mu}\nonumber
    \\
    &\qquad\times\bar{e}^{(i)a}_\alpha \bar{e}^{(l_2)d_2}_{\lambda_2}\hdots\bar{e}^{(l_D)d_D}_{\lambda_D}\Bigg] \; ,
\end{align}
where e.g. $(\delta g^{(i)2})^\alpha_{\;\mu}$ is shorthand for $\dg{i}{\alpha}{\lambda}\dg{i}{\lambda}{\mu}$.

Restricting to pairwise interactions where the DvN condition \eqref{SymmetricVierbeinCondition} holds, using the symmetry of the $\Tcoeffs$ coefficients, Eqs. \eqref{betas1} and \eqref{betas2} to relate these coefficients to the metric $\beta_m^{(i,j)}$, and Eqs. \eqref{ProductDeltas} and \eqref{VielbeinDeltas} to evaluate products of generalised deltas, one finds the following cubic potential:

\begin{widetext}\vspace{-1.5em}
\begin{align}
    &I_V^{(3)} = \sum_{i=0}^{N-1}\int \dd[D]x\sqrt{-\det\bar{g}_{(i)}}\nonumber
    \\
    &\times\Bigg\{-\frac{\kappa_i^3}{16}(\delta g^{(i)3}-4\delta g^{(i)2}\omega^{(i)}+16\delta g^{(i)}\omega^{(i)2}+32\omega^{(i)3})^\alpha_{\;\mu}\bar{W}^{(i)\mu}_{\;\;\;\;\;\;\alpha} \nonumber
    \\
    &\qquad + \frac{\kappa_i^3}{32}\tensor{\mathcal{A}}{^{(i)\mu}_{\alpha}^\nu_{\beta}}(\delta g^{(i)2}+8\delta g^{(i)}\omega^{(i)}-16\omega^{(i)2})^\alpha_{\;\mu}(\delta g^{(i)}-4\omega^{(i)})^\beta_{\;\nu} \nonumber
    \\
    &\qquad-\frac{\kappa_i^3}{48}\tensor{\mathcal{Z}}{^{(i)\mu}_\alpha^\nu_\beta^\rho_\gamma} (\delta g^{(i)}-4\omega^{(i)})^\alpha_{\;\mu}(\delta g^{(i)}-4\omega^{(i)})^\beta_{\;\nu}(\delta g^{(i)}-4\omega^{(i)})^\gamma_{\;\rho}\nonumber
    \\
    &\qquad + \sum_j\bigg[ \frac{\kappa_i^2\kappa_j}{16}[\mathcal{B}_{i,j}^{(+)}]^{\mu\;\;\nu}_{\;\;\alpha\;\;\beta}(\bar{S}_{i\rightarrow j})^\beta_{\;\sigma} (\delta g^{(i)2}+8\delta g^{(i)}\omega^{(i)}-16\omega^{(i)2})^\alpha_{\;\mu}(\delta g^{(j)}-4\omega^{(j)})^\sigma_{\;\nu}\nonumber
    \\
    &\qquad\qquad + \frac{\kappa_j^2\kappa_i}{16}[\mathcal{B}_{i,j}^{(+)}]^{\mu\;\;\nu}_{\;\;\alpha\;\;\beta} (\bar{S}_{i\rightarrow j})^\beta_{\;\sigma}(\delta g^{(j)2}+8\delta g^{(j)}\omega^{(j)}-16\omega^{(j)2})^\alpha_{\;\mu}(\delta g^{(i)}-4\omega^{(i)})^\sigma_{\;\nu}\nonumber
    \\
    &\qquad \qquad -\frac{\kappa_i^2\kappa_j}{16}[\mathcal{X}_{i,j}^{(+)}]^{\mu\;\;\nu\;\;\rho}_{\;\;\alpha\;\;\beta\;\;\gamma}(\bar{S}_{i\rightarrow j})^\gamma_{\;\sigma} (\delta g^{(i)}-4\omega^{(i)})^\alpha_{\;\mu}(\delta g^{(i)}-4\omega^{(i)})^\beta_{\;\nu}(\delta g^{(j)}-4\omega^{(j)})^\sigma_{\;\rho}\nonumber
    \\
    &\qquad\qquad - \frac{\kappa_j^2\kappa_i}{16}[\mathcal{Y}_{i,j}^{(+)}]^{\mu\;\;\nu\;\;\rho}_{\;\;\alpha\;\;\beta\;\;\gamma}(\bar{S}_{i\rightarrow j})^\beta_{\;\sigma}(\bar{S}_{i\rightarrow j})^\gamma_{\;\lambda}(\delta g^{(i)}-4\omega^{(i)})^\alpha_{\;\mu}(\delta g^{(j)}-4\omega^{(j)})^\sigma_{\;\nu}(\delta g^{(j)}-4\omega^{(j)})^\lambda_{\;\rho}\bigg]\nonumber
    \\
    &\qquad + \sum_k\bigg[\text{negative orientation terms with }j\rightarrow k \; ,\;\; \mathcal{B}^{(+)},\mathcal{X}^{(+)},\mathcal{Y}^{(+)}\rightarrow \mathcal{B}^{(-)},\mathcal{X}^{(-)},\mathcal{Y}^{(-)}\bigg]\Bigg\} \; ,\label{CubicPotential}
\end{align}
\end{widetext}
where $\mathcal{A}$ and $\mathcal{B}$ are as in Eqs. \eqref{Atens}--\eqref{Ctens}, and we have defined the new 6-index tensors:
\begingroup\allowdisplaybreaks
\begin{align}
         &\tensor{\mathcal{Z}}{^{(i)\mu}_\alpha^\nu_\beta^\rho_\gamma}\nonumber
         \\
         &=\sum_j\sum_{m=0}^D\frac{\beta_m^{(i,j)}}{m!}\delta^{\mu\nu\rho\lambda_1\hdots\lambda_m}_{\alpha\beta\gamma\gamma_1\hdots\gamma_m} (\bar{S}_{i\rightarrow j})^{\gamma_1}_{\;\lambda_1}\hdots(\bar{S}_{i\rightarrow j})^{\gamma_m}_{\;\lambda_m}\nonumber
         \\
         &\;+\sum_k\sum_{m=0}^D\frac{\beta_{D-m}^{(k,i)}}{m!}\delta^{\mu\nu\rho\lambda_1\hdots\lambda_m}_{\alpha\beta\gamma\gamma_1\hdots\gamma_m} (\bar{S}_{i\rightarrow k})^{\gamma_1}_{\;\lambda_1}\hdots(\bar{S}_{i\rightarrow k})^{\gamma_m}_{\;\lambda_m} \; ,
         \\
         &[\mathcal{X}_{i,j}^{(+)}]^{\mu\;\;\nu\;\;\rho}_{\;\;\alpha\;\;\beta\;\;\gamma}\nonumber
         \\
         &=\sum_{m=0}^D\frac{\beta_{m}^{(i,j)}}{(m-1)!}\delta^{\mu\nu\rho\lambda_1\hdots\lambda_{m-1}}_{\alpha\beta\gamma\gamma_1\hdots\gamma_{m-1}} 
         (\bar{S}_{i\rightarrow j})^{\gamma_1}_{\;\lambda_1}\hdots(\bar{S}_{i\rightarrow j})^{\gamma_{m-1}}_{\;\lambda_{m-1}} \; ,
         \\
         &[\mathcal{Y}_{i,j}^{(+)}]^{\mu\;\;\nu\;\;\rho}_{\;\;\alpha\;\;\beta\;\;\gamma}\nonumber
         \\
         &=\sum_{m=0}^D\frac{\beta_{m}^{(i,j)}}{(m-2)!}\delta^{\mu\nu\rho\lambda_1\hdots\lambda_{m-2}}_{\alpha\beta\gamma\gamma_1\hdots\gamma_{m-2}} (\bar{S}_{i\rightarrow j})^{\gamma_1}_{\;\lambda_1}\hdots(\bar{S}_{i\rightarrow j})^{\gamma_{m-2}}_{\;\lambda_{m-2}} \; ,
\end{align}\endgroup
as well as analogous negative orientation terms for $\mathcal{X}$ and $\mathcal{Y}$ given by the simultaneous exchanges $j\rightarrow k$, and $\beta_m\rightarrow\beta_{D-m}$. As in the quadratic case, $\mathcal{A}$ may be eliminated in favour of $\bar{W}$ and $\mathcal{B}$ using Eq. \eqref{ABW_relation}. A similar relation exists that will allow one to eliminate $\mathcal{Z}$ in favour of $\bar{W}$, $\mathcal{X}$ and $\mathcal{Y}$, namely\footnote{There are some interesting combinatorics at play here: schematically, at quadratic order, the relation \eqref{ABW_relation} is $\delta_{(2)} W = \mathcal{A} + \mathcal{B} S$, while at cubic order the relation \eqref{ZXY_relation} is $\delta_{(3)} W = \mathcal{Z} + 2\mathcal{X}S + \mathcal{Y}SS$; the coefficients of the terms on the right hand sides of these expressions form the second and third rows of Pascal's triangle! Presumably, at quartic order, there will then exist a similar relation involving 4 new 8-index tensor structures on the right hand side of $\delta_{(4)}W=(...)$ appearing with coefficients 1,3,3,1.}:
\begin{align}
    &\tensor{\mathcal{Z}}{^{(i)\mu}_\alpha^\nu_\beta^\rho_\gamma} \tensor{\chi}{^\beta_\nu}\tensor{\chi}{^\gamma_\rho} \nonumber
    \\
    &+ \sum_j \Bigg[2[\mathcal{X}_{i,j}^{(+)}]^{\mu\;\;\nu\;\;\rho}_{\;\;\alpha\;\;\beta\;\;\gamma}\tensor{\chi}{^\beta_\nu}\tensor{\chi}{^\sigma_\rho}(\bar{S}_{i\rightarrow j})^\gamma_{\;\sigma}\nonumber
    \\
    &\qquad\qquad+[\mathcal{Y}_{i,j}^{(+)}]^{\mu\;\;\nu\;\;\rho}_{\;\;\alpha\;\;\beta\;\;\gamma}\tensor{\chi}{^\sigma_\nu}\tensor{\chi}{^\lambda_\rho}(\bar{S}_{i\rightarrow j})^\beta_{\;\sigma}(\bar{S}_{i\rightarrow j})^\gamma_{\;\lambda}\nonumber
    \\
    &+ \sum_k \Bigg[2[\mathcal{X}_{i,k}^{(-)}]^{\mu\;\;\nu\;\;\rho}_{\;\;\alpha\;\;\beta\;\;\gamma}\tensor{\chi}{^\beta_\nu}\tensor{\chi}{^\sigma_\rho}(\bar{S}_{i\rightarrow k})^\gamma_{\;\sigma}\nonumber
    \\
    &\qquad\qquad+[\mathcal{Y}_{i,k}^{(-)}]^{\mu\;\;\nu\;\;\rho}_{\;\;\alpha\;\;\beta\;\;\gamma}\tensor{\chi}{^\sigma_\nu}\tensor{\chi}{^\lambda_\rho}(\bar{S}_{i\rightarrow k})^\beta_{\;\sigma}(\bar{S}_{i\rightarrow k})^\gamma_{\;\lambda}\nonumber\Bigg]
    \\
    &= \delta^{\mu\nu\rho}_{\lambda\beta\gamma} \tensor{\chi}{^\beta_\nu}\tensor{\chi}{^\gamma_\rho}\bar{W}^{(i)\lambda}_{\;\;\;\;\;\;\alpha} \; .\label{ZXY_relation}
\end{align}
Again the structure of the interaction terms ($\mathcal{X}$ and $\mathcal{Y}$) in terms of the generalised deltas ensures that $\delta g^{(i)0}_{\;\;\;\;\;0}$ will always appear in the action as a Lagrange multiplier, and $\delta g^{(i)j}_{\;\;\;\;\;0}$ will always appear as an auxiliary field, leading to the necessary constraints that guarantee the removal of the BD ghost in the standard way.

One can check that around a proportional background Eq. \eqref{CubicPotential} reduces to the cubic order potential given in \cite{heavy_spin2_DM}. We recall that when $\gi{i}=a_i^2\bar{g}_{\mu\nu}$, one has $(\bar{S}_{i\rightarrow j})^\mu_{\;\nu}=(a_j/a_i)\delta^\mu_\nu$, so the Lorentz constraints become trivial, as $\delta g$ and $\bar{S}$ commute. Therefore, around proportional backgrounds, all of the local Lorentz fields drop out of the cubic potential, and we only need to care about the terms involving exclusively metric perturbations. Using Eq. \eqref{ProductDeltas} to evaluate products of deltas, alongside Eqs. \eqref{PropB1} and \eqref{PropB2} for the $\mathcal{B}$-tensors, one finds that the relevant new tensor structures dictating the form of the spin-2 interactions are:
\begingroup\allowdisplaybreaks\vspace{-0.5em}
\begin{align}
    [\mathcal{X}_{i,j}^{(+)}]^{\mu\;\;\nu\;\;\rho}_{\;\;\alpha\;\;\beta\;\;\gamma} &= \eta_{i,j}^{(+)}\delta^{\mu\nu\rho}_{\alpha\beta\gamma}  \; ,
    \\[0.3em]
    [\mathcal{Y}_{i,j}^{(+)}]^{\mu\;\;\nu\;\;\rho}_{\;\;\alpha\;\;\beta\;\;\gamma} &= \frac{a_i}{a_j}\left[\sigma_{i,j}^{(+)}-\eta_{i,j}^{(+)}\right] \delta^{\mu\nu\rho}_{\alpha\beta\gamma}  \; ,
    \\[0.3em]
    [\mathcal{X}_{i,k}^{(-)}]^{\mu\;\;\nu\;\;\rho}_{\;\;\alpha\;\;\beta\;\;\gamma} &= \eta_{i,k}^{(-)}\delta^{\mu\nu\rho}_{\alpha\beta\gamma}  \; ,
    \\[0.3em]
    [\mathcal{Y}_{i,k}^{(-)}]^{\mu\;\;\nu\;\;\rho}_{\;\;\alpha\;\;\beta\;\;\gamma} &= \frac{a_i}{a_k}\left[\sigma_{i,k}^{(-)}-\eta_{i,k}^{(-)}\right] \delta^{\mu\nu\rho}_{\alpha\beta\gamma}  \; ,
\end{align}\endgroup
where we have defined:
\begin{align}
    \eta_{i,j}^{(+)} = \sum_{m=0}^D \beta_m^{(i,j)}\binom{D-3}{m-1} \left(\frac{a_j}{a_i}\right)^{m-1} \; ,
    \\
    \eta_{i,k}^{(-)} = \sum_{m=0}^D \beta_{D-m}^{(k,i)}\binom{D-3}{m-1} \left(\frac{a_k}{a_i}\right)^{m-1} \; ,
\end{align}
related by:
\begin{equation}
    \eta_{i,j}^{(+)} = \sigma_{i,j}^{(+)} - \left(\frac{a_j}{a_i}\right)^{D-2}\eta_{j,i}^{(-)}  \; .
\end{equation}
Using Eqs. \eqref{ABW_relation} and \eqref{ZXY_relation} to eliminate $\mathcal{A}$ and $\mathcal{Z}$, and ignoring the terms left multiplying $\bar{W}$ (which are just contributions from the background curvature), one finds that the spin-2 interaction potential at cubic order reads:

\begin{widetext}\vspace{-2em}
    \begin{align}
        &I_V^{(3)} \supset \sum_{i=0}^{N-1} \int \dd[D]x\sqrt{-\det \bar{g}_{(i)}}
        \nonumber
        \\
        &\times\Bigg\{ \frac{\kappa_i^3}{48}\left( \delta g^{(i)3} - 3\delta g^{(i)} \dg{i}{\mu}{\nu}\dg{i}{\nu}{\mu} +2\dg{i}{\mu}{\nu}\dg{i}{\nu}{\rho}\dg{i}{\rho}{\mu}\right)\left[\sum_j\frac{a_j}{a_i}\eta_{i,j}^{(+)} + \sum_k \frac{a_k}{a_i} \eta_{i,k}^{(-)}\right] 
        \nonumber
        \\
        &+ \frac{\kappa_i^3}{96}\left(2\delta g^{(i)3} - 9 \delta g^{(i)}\dg{i}{\mu}{\nu}\dg{i}{\nu}{\mu} + 7\dg{i}{\mu}{\nu}\dg{i}{\nu}{\rho}\dg{i}{\rho}{\mu}\right)\left[\sum_j \frac{a_j}{a_i}\sigma_{i,j}^{(+)} + \sum_k \frac{a_k}{a_i}\sigma_{i,k}^{(-)}\right]
        \nonumber
        \\
        &-\frac{1}{16}\Bigg[\sum_j \frac{a_j}{a_i}\eta_{i,j}^{(+)}\bigg[\kappa_i^2\kappa_j\left(\delta g^{(i)2}\delta g^{(j)} - \delta g^{(j)}\dg{i}{\mu}{\nu}\dg{i}{\nu}{\mu} - 2\delta g^{(i)}\dg{i}{\mu}{\nu}\dg{j}{\nu}{\mu} + 2\dg{i}{\mu}{\nu}\dg{i}{\nu}{\rho}\dg{j}{\rho}{\mu} \right) 
        \nonumber
        \\
        &\qquad\qquad\qquad\qquad - \kappa_i\kappa_j^2\left(\delta g^{(j)2}\delta g^{(i)} - \delta g^{(i)}\dg{j}{\mu}{\nu}\dg{j}{\nu}{\mu} - 2\delta g^{(j)}\dg{j}{\mu}{\nu}\dg{i}{\nu}{\mu} + 2\dg{j}{\mu}{\nu}\dg{j}{\nu}{\rho}\dg{i}{\rho}{\mu}\right)\bigg]
        \nonumber
        \\
        &\qquad + \sum_k \frac{a_k}{a_i}\eta_{i,k}^{(-)}\bigg[\kappa_i^2\kappa_k\left(\delta g^{(i)2}\delta g^{(k)} - \delta g^{(k)}\dg{i}{\mu}{\nu}\dg{i}{\nu}{\mu} - 2\delta g^{(i)}\dg{i}{\mu}{\nu}\dg{k}{\nu}{\mu} + 2\dg{i}{\mu}{\nu}\dg{i}{\nu}{\rho}\dg{k}{\rho}{\mu} \right) 
        \nonumber
        \\
        &\qquad\qquad\qquad\qquad - \kappa_i\kappa_k^2\left(\delta g^{(k)2}\delta g^{(i)} - \delta g^{(i)}\dg{k}{\mu}{\nu}\dg{k}{\nu}{\mu} - 2\delta g^{(k)}\dg{k}{\mu}{\nu}\dg{i}{\nu}{\mu} + 2\dg{k}{\mu}{\nu}\dg{k}{\nu}{\rho}\dg{i}{\rho}{\mu}\right)\bigg]
        \nonumber \Bigg]
        \\
        &+ \frac{1}{16}\Bigg[\sum_j \frac{a_j}{a_i}\sigma_{i,j}^{(+)}\bigg[\kappa_i^2\kappa_j \left(\delta g^{(j)}\dg{i}{\mu}{\nu}\dg{i}{\nu}{\mu}-\dg{i}{\mu}{\nu}\dg{i}{\nu}{\rho}\dg{j}{\rho}{\mu}\right)
        \nonumber
        \\
        &\qquad\qquad\qquad\qquad+\kappa_i\kappa_j^2\left(2\delta g^{(i)}\dg{j}{\mu}{\nu}\dg{j}{\nu}{\mu} - \delta g^{(j)2}\delta g^{(i)} - 3\dg{j}{\mu}{\nu}\dg{j}{\nu}{\rho}\dg{i}{\rho}{\mu} + 2\delta g^{(j)}\dg{j}{\mu}{\nu}\dg{i}{\nu}{\mu}\right)\bigg]
        \nonumber
        \\
        &\qquad + \sum_k \frac{a_k}{a_i}\sigma_{i,k}^{(-)}\bigg[\kappa_i^2\kappa_k \left(\delta g^{(k)}\dg{i}{\mu}{\nu}\dg{i}{\nu}{\mu}-\dg{i}{\mu}{\nu}\dg{i}{\nu}{\rho}\dg{k}{\rho}{\mu}\right)
        \nonumber
        \\
        &\qquad\qquad\qquad\qquad+\kappa_i\kappa_k^2\left(2\delta g^{(i)}\dg{k}{\mu}{\nu}\dg{k}{\nu}{\mu} - \delta g^{(k)2}\delta g^{(i)} - 3\dg{k}{\mu}{\nu}\dg{k}{\nu}{\rho}\dg{i}{\rho}{\mu} + 2\delta g^{(k)}\dg{k}{\mu}{\nu}\dg{i}{\nu}{\mu}\right)\bigg]\Bigg]\Bigg\} \; .\label{CubicPotentialProp}
    \end{align}
\end{widetext}
This expression is quite horrendous, but one can check that upon redefining $\delta\gi{i}\rightarrow \delta\gi{i}/\kappa_i$ to match the normalisation used in \cite{heavy_spin2_DM}, working explicitly in $D=4$ spacetime dimensions, setting all $a_i=1$ to match the background they expanded around, and limiting oneself to only $N=2$ interacting metrics (they denote $\delta\gi{1}\equiv h_{\mu\nu}$ and $\delta\gi{2}\equiv l_{\mu\nu}$, with the interaction positively oriented from $h$ to $l$), Eq. \eqref{CubicPotentialProp} agrees precisely with their cubic potential for bigravity (their Eq. B.10).

\section{Conclusion}\label{sec:conclusion}

In this work, we have developed a new procedure for computing the structure of metric perturbations around arbitrary background spacetimes in dRGT-type multi-gravity theories. Our approach hinges on the equivalence between the metric and vielbein formalisms of multi-gravity in theories involving exclusively pairwise interactions, as performing the initial perturbation of the action in vielbein form circumvents the need to deal with the cumbersome matrix square root underpinning the interactions in the metric formalism. This is the first real advantage of our approach over previous approaches. The second is that expressing the perturbed interaction terms via the generalised Kronecker delta elucidates their ghost-free nature in a much more transparent manner, as it becomes clear at the level of the action that the ghost-killing constraint will always be present, at any order, regardless of the background spacetime. The final and most important advantage is the generality of our approach: it works very naturally in arbitrary spacetime dimension, for an arbitrary number of interacting metrics, and crucially, to an arbitrary order in perturbation theory, where previous approaches were limited to only linear order.

To verify that our formalism functions as intended, we used it to reproduce the linearised field equations of multi-gravity around three commonly occurring example background spacetimes -- proportional, cosmological and black hole -- in the process also extending some of these results to multiple interacting metrics, and in the black hole case, providing for the first time the structure of the linearised mass term around a non-proportional, \emph{rotating} black hole. To then demonstrate the power of our formalism beyond just the linear order, we computed the cubic order multi-gravity potential around an arbitrary background, which to our knowledge constitutes the first time this has been done. We verified that our generic cubic potential reduces to the only known cubic potential from bigravity \cite{heavy_spin2_DM} around a proportional background.

This work should prove useful to anybody who wishes to study perturbations in massive gravity theories around complicated background spacetimes and/or to higher than linear order. For example, we mentioned one potential future use earlier: some (non-proportional) black hole solutions in multi-gravity are expected to be pathological, but the pathology likely only surfaces nonlinearly; using our formalism to determine the perturbation structure around such solutions to cubic order should hopefully shed some light on this issue. Another example from the realm of cosmology is that by going to higher orders in cosmological perturbation theory one may begin to study, for instance, scalar-induced gravitational waves in multi-gravity theories, where the nonlinear interactions between the scalar and tensor metric perturbations that are decoupled at linear level become important, with potentially measurable physical effects. These are just two examples, but the possibilities are of course many. We hope our formalism provides a helpful tool to those wishing to study such interesting questions in the future.

\section*{Acknowledgements}

I would like to thank Tasos Avgostidis and Paul Saffin for useful comments, and especially Joakim Flinckman for pointing out an important mistake I had made in a previous version of this manuscript. I acknowledge support from a UK Science and Technology Facilities Council studentship, and from David Stefanyszyn's Nottingham Research Fellowship from the University of Nottingham. Calculations involving tensor contractions around cosmological and black hole backgrounds were aided by use of the \emph{xAct} Mathematica package suite \cite{xAct}. For the purpose of open access, I have applied a Creative Commons Attribution (CC BY) licence to any Author Accepted Manuscript version arising.

\section*{Data Access Statement}

No new data were created or analysed in this study.

\appendix
\section{Details of the previous approaches to linear multi-gravity perturbations}\label{app:old perts}

In this appendix we provide some details of the previous approaches to multi-gravity perturbations in the metric formalism around arbitrary backgrounds, following \cite{MG_arbitrary_BG,Syzygies,linear_spin2_gen_BG} and later \cite{Volkov_arbitraryBG,Volkov_arbitraryBG_detailed}. We will explain where the complications arise that motivated our alternate procedure.

\subsection{Matrix equation approach}

As mentioned in footnote \ref{footnote_matrix}, in matrix form, the $W$-tensor at background level is typically written as:
\begin{equation}\label{W_matrix}
\begin{split}
    \Wi{i} &= \sum_j\sum_{m=0}^D (-1)^m \beta_m^{(i,j)} Y^\mu_{(m)\nu}(S_{i\rightarrow j})
    \\
    &+ \sum_k\sum_{m=0}^D (-1)^m \beta_{D-m}^{(k,i)} Y^\mu_{(m)\nu}(S_{i\rightarrow k})\; ,
\end{split}
\end{equation}
where the matrices $Y_{(m)}(S)$ are defined by:
\begin{equation}
    Y_{(m)}(S) = \sum_{n=0}^m(-1)^n S^{m-n}e_n(S) \; ,
\end{equation}
with the elementary symmetric polynomials given iteratively in terms of the matrix traces as:
\begin{equation}\label{sympols_matrix}
    e_n(S) = -\frac{1}{n}\sum_{m=1}^n(-1)^m\Tr(S^m)e_{n-m}(S) \; ,
\end{equation}
starting from $e_0(S)=1$. Eq. \eqref{W_matrix} is equivalent to our Eq. \eqref{W_metric}, and Eq. \eqref{sympols_matrix} is equivalent to our Eq. \eqref{sym pols}.

To derive the linearised field equations using the approach of \cite{MG_arbitrary_BG,Syzygies}, one considers the first order variation of the $W$-tensor in matrix form:
\begin{equation}\label{delta W}
\begin{split}
        \delta\Wi{i} &= \delta g^{(i)\mu}_{\;\;\;\;\;\lambda}\bar{W}^{(i)\lambda}_{\;\;\;\;\;\;\nu} 
        \\
        &+ \sum_j \sum_{m=0}^D(-1)^m\beta_m^{(i,j)}\delta Y_{(m)\nu}^\mu(S_{i\rightarrow j})
        \\
        &+ \sum_k\sum_{m=0}^D(-1)^m\beta_{D-m}^{(k,i)}\delta Y_{(m)\nu}^\mu(S_{i\rightarrow k}) \; .
\end{split}
\end{equation}
The variation of the $Y$'s is given (for any given $i\rightarrow j$ interaction) by:
\begin{align}\label{delta Y}
    \delta Y_{(m)}(S) = \sum_{k=1}^m (-1)^k &\Bigg[ S^{m-k}\delta e_k(S)
    \\
    &-e_{k-1}(S)\sum_{n=0}^{m-k} S^n \delta S S^{m-k-n} \Bigg] \; ,\nonumber
\end{align}
where the variation of the symmetric polynomials is found from Eq. \eqref{sympols_matrix} to be:
\begin{equation}\label{delta ek}
    \delta e_k(S) = -\sum_{n=1}^k (-1)^n\Tr(S^{n-1}\delta S) e_{k-n}(S) \; .
\end{equation}

The complication lies in the fact that $\delta S$ is given by the \emph{matrix} equation:
\begin{equation}\label{Sylvester eq}
    S\delta S + \delta S S = \delta S^2 \; ,
\end{equation}
and therefore determining its form is not as straightforward as simply starting from $S^2_{i\rightarrow j}=g^{-1}_{(i)}g_{(j)}$ and Taylor expanding the square root (although it is possible to do this when $S\propto\mathbbm{1}$, which \emph{is} the case for the proportional solutions, and is the reason that they are easier to deal with). Around a generic background solution, this is a \emph{Sylvester matrix equation}, of the form:
\begin{equation}
    AX-XB=C \; ,
\end{equation}
where $A$, $B$ and $C$ are given constant matrices and one wishes to solve for the unknown matrix $X$. The solution to the Sylvester equation is known in the mathematical literature, and is given by the following expression \cite{Sylvester_matrix_eq}:
\begin{equation}
    X = q_B^{-1}(A)\sum_{k=1}^D\sum_{n=0}^{k-1}(-1)^k e_{D-k}(B)A^{k-n-1}CB^n \; ,
\end{equation}
where $q_B(A)$ is the unique polynomial in the matrix $A$ whose coefficients are the same as those of the characteristic polynomial of $B$ ($q_B^{-1}(A)$ is then the inverse of this matrix); that is:
\begin{equation}
    q_B(A) = \sum_{m=0}^D (-1)^m e_{D-m}(B) A^m \; .
\end{equation}

In our case, comparison with Eq. \eqref{Sylvester eq} tells us that we have $A=S$, $B=-S$ and $C=\delta S^2$. Therefore, the solution for $\delta S$ is:
\begin{equation}\label{delta S}
    \delta S = q^{-1}_{-S}(S)\sum_{k=1}^D\sum_{n=0}^{k-1}(-1)^{n+k} e_{D-k}(-S)S^{k-n-1}\delta S^2 S^n \; .
\end{equation}
One can easily obtain $\delta S^2_{i\rightarrow j}$ in terms of either the metric perturbations of $g_{(i)}$ and $g_{(j)}$, or of their inverses, by starting from $S^2_{i\rightarrow j}=g^{-1}_{(i)}g_{(j)}$ and substituting in $g_{(i)}=\bar{g}_{(i)}+\delta g_{(i)}$ for the perturbed metrics. The result is:
\begin{align}
    \delta S^2_{i\rightarrow j} &= g_{(i)}^{-1}\left[\delta g_{(j)}-\delta g_{(i)}S^2_{i\rightarrow j}\right] 
    \\
    &= \left[S^2_{i\rightarrow j}\delta g_{(j)}^{-1}-\delta g_{(i)}^{-1}\right]g_{(j)} \; ,
\end{align}
or in components:
\begin{align}
    (\delta S^2_{i\rightarrow j})^\mu_{\;\nu} &= g_{(i)}^{\mu\lambda}\left[\delta g_{(j)\lambda\nu}-\delta g_{(i)\lambda\sigma}(S^2_{i\rightarrow j})^\sigma_{\;\nu}\right] 
    \\
    &= \left[(S^2_{i\rightarrow j})^\mu_{\;\lambda}\delta g_{(j)}^{\lambda\sigma}-\delta g_{(i)}^{\mu\sigma}\right]g_{(j)\sigma\nu} \; .
\end{align}

Substituting either of these expressions into Eq. \eqref{delta S} determines $\delta S$, which one can then substitute into Eq. \eqref{delta Y} to get the $Y$ variations, and lastly substitute these into Eq. \eqref{delta W} to determine the linearised $W$-tensors. However, the difficulty with this approach is that inverting the matrix $q_{-S}(S)$ can become quite a challenge around complicated backgrounds. This motivated the authors of \cite{MG_arbitrary_BG,Syzygies} to develop a second approach in \cite{linear_spin2_gen_BG}.

\subsection{Redefined fluctuation variables approach}

In bigravity, where there are just two metrics, $\gi{1}\equiv g_{\mu\nu}$ and $\gi{2}\equiv f_{\mu\nu}$, and a single interaction $S_{g\rightarrow f}\equiv S$ (so $S_{f\rightarrow g}=S^{-1}$), the necessity to solve the Sylvester matrix equation can be circumvented by redefining the metric perturbations in a clever way. It was shown in \cite{linear_spin2_gen_BG} that by absorbing $S$ into the definition of the fluctuation variables as:
\begin{align}
    \delta g_{\mu\nu} &= 2\delta^\lambda_{(\mu}S^\sigma_{\;\nu)}\delta g'_{\sigma\lambda} \; ,\label{delta g'}
    \\
    \delta f_{\mu\nu} &= 2\delta^\lambda_{(\mu}[S^{-1}]^\sigma_{\;\nu)}\delta f'_{\sigma\lambda} \; ,\label{delta f'}
\end{align}
then so long as $S$ and $-S$ do not share any common eigenvalues (i.e. no zeroes), $\delta S$ and $\delta S^{-1}$ can be uniquely determined in terms of $\delta g'$ and $\delta f'$ as:
\begin{align}
    \delta S &= -g^{-1}\delta g' S^2 + S^{-1} g^{-1}\delta f' S^{-1} \; ,
    \\
    \delta S^{-1} &= -f^{-1} \delta f' S^{-2} + S f^{-1} \delta g' S \; ,
\end{align}
symmetric under $g\leftrightarrow f$ exchange. These expressions are significantly simpler than Eq. \eqref{delta S}, and indeed the authors used them to demonstrate the existence of the ghost-killing constraint at the level of the linearised field equations in a covariant manner.

However, there are two problems with this approach, as alluded to in the introduction. The first is that by making the field redefinitions \eqref{delta g'} and \eqref{delta f'}, the kinetic part of the field equations now contains derivatives acting on $S$ and $S^{-1}$, which is undesirable, but ultimately not catastrophic. The real problem is that this procedure does not work as soon as the theory contains additional interacting metrics i.e. once one begins to consider true multi-gravity theories. For example, suppose $g_{\mu\nu}$ interacts with both $f_{\mu\nu}$ and an additional metric $h_{\mu\nu}$; by redefining the perturbations of $g_{\mu\nu}$ and $f_{\mu\nu}$ via Eqs. \eqref{delta g'} and \eqref{delta f'} to determine $\delta S_{g\rightarrow f}$, one loses the ability to do the same thing for $S_{g\rightarrow h}$ because the perturbations of $g_{\mu\nu}$ have already been redefined -- it is only ever possible to absorb \emph{one} interaction into the field redefinitions. Thus, in multi-gravity theories, up until now one was forced to succumb to solving the Sylvester matrix equation for each $\delta S_{i\rightarrow j}$, which was not ideal. The improved procedure we have developed in this paper, however, works for any number of metrics, does not require any such field redefinitions to be made, and trades the complicated matrix equations for the algebraic Lorentz constraints.

\subsection{Mazuet and Volkov's vielbein approach}\label{app:Volkov}

As discussed in the introduction, the approach from \cite{Volkov_arbitraryBG,Volkov_arbitraryBG_detailed}, like ours, is able to circumvent the difficulties associated with the square root matrices by working instead with the vielbein perturbations. However, their work was applied specifically to 4-dimensional dRGT gravity, with fixed reference metric (where equivalence between metric and vielbein formulations is automatic as there is only one interaction), rather than a generic multi-gravity theory; we will see that many of their calculations become challenging practically once more dynamical metrics or extra dimensions are involved.

In the notation of \cite{Volkov_arbitraryBG,Volkov_arbitraryBG_detailed}, there are again two metrics: the physical metric $\gi{1}\equiv g_{\mu\nu}$, with associated vierbein $\vbein{1}{\mu}{a}\equiv e_\mu^{\;a}$, and the fixed reference metric $\gi{2}\equiv f_{\mu\nu}$, with associated vierbein $\vbein{2}{\mu}{a}\equiv\phi_\mu^{\;a}$. They write the background field equations for the physical metric explicitly in terms of traces and powers of a matrix they call $\gamma$, defined by:
\begin{equation}\label{gamma_defs}
    \gamma^\mu_{\;\nu} = e^\mu_{\;a} \phi_\nu^{\;a} \; ,
\end{equation}
in the following manner (Eq. 3.30 in \cite{Volkov_arbitraryBG_detailed}):
\begin{equation}\label{Volkov_fieldeqs}
\begin{split}
    G_{\mu\nu} &+ \beta_0 g_{\mu\nu} 
    \\
    &+ \beta_1 \left[e_1(\gamma)g_{\mu\nu} - \gamma_{\mu\nu}\right] 
    \\
    &+ \beta_2 \left[\gamma^2_{\mu\nu} - e_1(\gamma)\gamma_{\mu\nu} + e_2(\gamma) g_{\mu\nu}\right]
    \\
    &+ \beta_3 \left[-\gamma^3_{\mu\nu} + e_1(\gamma)\gamma^2_{\mu\nu} - e_2(\gamma)\gamma_{\mu\nu}+e_3(\gamma)g_{\mu\nu}\right] = 0 \; .
\end{split}
\end{equation}
Of course, the matrix $\gamma$ is nothing more than $S_{g\rightarrow f}$ (c.f. Eq. \eqref{SijVielbein}), and consequently the field equations as written above are simply in the form of our Eq. \eqref{W_matrix}.

They linearise Eq. \eqref{Volkov_fieldeqs} by perturbing the vierbein of the physical metric\footnote{The reference vierbein, $\phi_\mu^{\;a}$, is not perturbed, as it is completely fixed in dRGT massive gravity.} as $e_\mu^{\;a}\rightarrow \bar{e}_\mu^{\;a}+\delta e_\mu^{\;a}$, then computing the variation of the matrix $\gamma$ in terms of the vierbein perturbations, which in principle contain all 16 components -- both physical and non-physical. The authors of \cite{Volkov_arbitraryBG,Volkov_arbitraryBG_detailed} do \emph{not} explicitly separate the vierbein perturbations into metric and local Lorentz degrees of freedom as we do in our Eq. \eqref{VielbeinPerturbations}. Instead, they elect to simply project the vierbein perturbations onto the background vierbein, resulting in a description where all 16 degrees of freedom contained within $\delta e_\mu^{\;a}$ are carried by the following non-symmetric tensor:
\begin{equation}
    X^\mu_{\;\nu} \equiv \bar{e}^\mu_{\;a}\delta e_\nu^{\;a} \; .
\end{equation}
A benefit of our approach is that, thanks to Eq. \eqref{VielbeinPerturbations}, one is able to explicitly keep track of which degrees of freedom within $X_{\mu\nu}$ are unphysical---one may identify (c.f. Eqs. \eqref{1storder}--\eqref{3rdorder}):
\begin{align}
    \tensor{X}{^{(1)\mu}_\nu} &= \frac{\kappa}{2}(\delta g - 4\omega)^\mu_{\;\nu} \; ,
    \\
    \tensor{X}{^{(2)\mu}_\nu} &= -\frac{\kappa^2}{8}(\delta g^2 -8 \omega\delta g  + 16\omega^2)^\mu_{\;\nu} \; ,
    \\
    \tensor{X}{^{(3)\mu}_\nu} &= \frac{\kappa^3}{16}(\delta g^3 - 4\delta g^2\omega + 16\omega^2\delta g + 32\omega^3 )^\mu_{\;\nu} \; ,
    \\
    \vdots\nonumber
\end{align}
from which one can clearly see that e.g. $\kappa\delta g_{\mu\nu}=2X^{(1)}_{(\mu\nu)}$, as in Eq. 3.22 of \cite{Volkov_arbitraryBG_detailed}.

To linear order, directly from Eq. \eqref{gamma_defs}, one finds that, in terms of the tensor $X_{\mu\nu}$, the variation of $\gamma^\mu_{\;\nu}$ (which is really $S^\mu_{\;\nu}$) is\footnote{If the reference metric were dynamical, Eq. \eqref{gamma_perts} would instead read $\delta \gamma^\mu_{\;\nu} = - \bar{\gamma}^\lambda_{\;\nu}\tensor{X}{^{(g)\mu}_\lambda} + \bar{\gamma}^\mu_{\;\lambda}\tensor{X}{^{(f)\lambda}_\nu}$, which is why the difference expressions $[X]_{i,j}$ (c.f. Eq. \eqref{DiffMetPerts}) show up in our effective mass terms from section \ref{sec:perts}.}:
\begin{equation}\label{gamma_perts}
    \delta\gamma^\mu_{\;\nu} =  - \bar{\gamma}^\lambda_{\;\nu}\tensor{X}{^{\mu}_\lambda} \; .
\end{equation}
The effective mass term, Eq. 3.26 in \cite{Volkov_arbitraryBG_detailed}, is then computed by the direct substitution of Eq. \eqref{gamma_perts} into Eq. \eqref{Volkov_fieldeqs}:
\begin{equation}\label{Volkov_massmatrix}
    \begin{split}
        \mathcal{M}_{\mu\nu} &= \beta_1 \left[\bar{\gamma}^\lambda_{\;\mu} X_{\lambda\nu} - g_{\mu\nu} \bar{\gamma}_{\alpha\beta}X^{\alpha\beta}\right]
        \\
        &+ \beta_2 \big[-\bar{\gamma}^\alpha_{\;\mu}\bar{\gamma}^\beta_{\;\nu}X_{\alpha\beta} - (\bar{\gamma}^2)^\lambda_{\;\mu}X_{\lambda\nu} + \bar{\gamma}_{\mu\nu}\bar{\gamma}_{\alpha\beta}X^{\alpha\beta}
        \\
        &\qquad + \bar{\gamma} \bar{\gamma}^\lambda_{\;\mu}X_{\lambda\nu} + g_{\mu\nu}\left((\bar{\gamma}^2)_{\alpha\beta}X^{\alpha\beta} - \bar{\gamma} \bar{\gamma}_{\alpha\beta}X^{\alpha\beta}\right)\big]
        \\
        &+ \beta_3 \det\bar{\gamma} \left[X_{\mu\lambda} (\bar{\gamma}^{-1})^\lambda_{\;\nu} - X (\bar{\gamma}^{-1})_{\mu\nu}\right] \; .
    \end{split}
\end{equation}
One can check that this equation agrees precisely with the $\mathcal{B}$-tensor terms from our Eq. \eqref{GeneralLinearisedFieldEqs}, upon expanding out the generalised deltas (the $\beta_3$ term also requires use of the Cayley-Hamilton relation -- Eq. 3.10 in \cite{Volkov_arbitraryBG_detailed} -- to replace the dependence on $\bar{\gamma}^{-1}$ with a dependence on $\bar{\gamma}$ itself). However, the structure of each of the terms is arguably clearer in our expressions, holds for multiple dynamical metrics, and easily extends to general spacetime dimension, where this would likely prove a small challenge if using their approach.

For completeness, the antisymmetric part of their linearised field equations leads to $\mathcal{M}_{[\mu\nu]}=0$ as we expect; taking Eq. \eqref{Volkov_massmatrix} for the effective mass term, this implies the following 6 constraint equations:
\begin{equation}
    \bar{\gamma}^\lambda_{\;\mu}X_{\lambda\nu} = \bar{\gamma}^\lambda_{\;\nu}X_{\lambda\mu} \; ,
\end{equation}
which form Eq. 4.3 in \cite{Volkov_arbitraryBG_detailed}, and are of course equivalent to our linearised Lorentz constraints \eqref{Linearised_Lorentz_constraints}. They show explicitly that 5 more constraints arise from the linearised field equations: 4 from taking their divergence, and one extra from taking an additional divergence then subtracting off their trace. In total, there are hence 11 constraints; consequently, the tensor $X_{\mu\nu}$ propagates 5 physical degrees of freedom, as befitting a massive spin-2 field in 4 dimensions.

Finally, we would like to comment on one additional interesting thing the authors of \cite{Volkov_arbitraryBG,Volkov_arbitraryBG_detailed} do, which works well in dRGT gravity but is much more difficult (and less useful) in multi-gravity. After computing the effective mass term \eqref{Volkov_massmatrix}, the authors decide to attempt to remove all dependence of the linearised field equations on the matrix $\bar{\gamma}_{\mu\nu}$, instead expressing them only in terms of the tensor $X_{\mu\nu}$ and powers of the background curvature of the physical metric. To do so, they treat the background field equations \eqref{Volkov_fieldeqs} as algebraic, polynomial equations for $\bar{\gamma}_{\mu\nu}$ that may be solved in terms of $g_{\mu\nu}$ and $R_{\mu\nu}$. Since they work in $D=4$, these equations are cubic, so the solutions have the structure:
\begin{equation}\label{gammasol}
    \bar{\gamma}_{\mu\nu} = y_0 g_{\mu\nu} + y_1 R_{\mu\nu} + y_2 R^2_{\mu\nu} + y_3 R^3_{\mu\nu} \; ,
\end{equation}
for some set of coefficients $y_m$, which can (in principle) be solved for by injecting the above ansatz into Eqs. \eqref{Volkov_fieldeqs}. Upon solving for the $y_m$, one can also (in principle) determine the effective mass term as a function of $g_{\mu\nu}$ and $R_{\mu\nu}$ by substituting the solution for $\bar{\gamma}_{\mu\nu}$ into Eq. \eqref{Volkov_massmatrix}; one finds a similar structure:
\begin{equation}
    \mathcal{M}_{\mu\nu} = B_0 g_{\mu\nu} + B_1 R_{\mu\nu} + B_2 R_{\mu\nu}^2 + B_3 R_{\mu\nu}^3 \; ,
\end{equation}
for some new coefficients $B_m$. 

In practice, this process can be debilitatingly difficult for a massive gravity theory with generic $\beta_m$. Indeed, it was only done explicitly in \cite{Volkov_arbitraryBG_detailed} for two simplified models: one containing only non-vanishing $\beta_1$, and the other containing only non-vanishing $\beta_3$. As before, the procedure only gets more difficult in higher dimensions, as the order of the polynomial equation the one must solve for $\bar{\gamma}_{\mu\nu}$ (and hence the highest power of $R_{\mu\nu}$ in the above expressions) increases. Furthermore, it only really makes sense to do this at all for dRGT gravity, where the reference metric is a fixed input to the theory: by expressing $\bar{\gamma}_{\mu\nu}$ in the form suggested by Eq. \eqref{gammasol}, one is really tuning the reference metric to ensure that an arbitrary $g_{\mu\nu}$ will always satisfy its background field equations; when the reference metric has its own dynamics, there is no guarantee that this tuning will satisfy both sets of field equations simultaneously. In this case, if one wished to proceed with their approach, one would have to solve both sets of background field equations simultaneously, leading to an expression for $\bar{\gamma}_{\mu\nu}$ given in terms of the background curvatures of \emph{both} metrics. This represents a significant increase in complexity, which of course only gets worse when there are additional dynamical metrics and hence more simultaneous equations. Instead, we argue, when one is working with a genuine multi-gravity theory, one should always first solve the background field equations to determine the form of the various background metrics. Once this is done, all of the background $\bar{S}_{i\rightarrow j}$ matrices will be known (provided that one is able to compute the matrix square roots, or equally, the associated vielbeins and their inverses); it then becomes a relatively simple task to substitute these matrices into Eq. \eqref{GeneralLinearisedFieldEqs} to find the structure of the effective mass terms/Eq. \eqref{Linearised_Lorentz_constraints} to determine the linearised Lorentz constraints.



\section{Recovery of non-proportional Schwarzschild perturbation structure from our general black hole perturbation expressions}\label{app:BH}

In this appendix we demonstrate how to recover the perturbation structure around non-proportional, non-rotating black holes in 4-dimensional multi-gravity, given in \cite{GL_unified,Stability_nonbidiag,BH2}, from our general black hole perturbation expressions \eqref{BHperts1} and \eqref{BHperts2}.

To start, we need to expand a little bit on how multi-metric black hole solutions are actually constructed. Let us assume that we are in vacuum, so the black holes are uncharged. With the ansatz \eqref{KdSMultiGrav} for the metrics, the Einstein tensors are simply:
\begin{equation}\label{EinsteinTensorBHs}
    \Gi{i} = -\frac{\Lambda}{a_i^2}\delta^\mu_\nu \; ,
\end{equation}
irrespective of the scalar functions $\phi_i$, while the $W$-tensors have components:
\begin{align}
    \Wi{i} &= \delta^\mu_\nu\Bigg[\sum_j\sum_{m=0}^{D} \beta_m^{(i,j)} \binom{D-1}{m}\left(\frac{a_j}{a_i}\right)^m \nonumber
    \\
    &\qquad+ \sum_k\sum_{m=0}^{D} \beta_{D-m}^{(k,i)} \binom{D-1}{m} \left(\frac{a_k}{a_i}\right)^m \Bigg]\nonumber
    \\
    &+l^\mu l_\nu\Bigg[\sum_j\frac{a_{j}}{a_i}\left(\phi_i-\phi_j\right)\sigma_{i,j}^{(+)} \nonumber
    \\
    &\qquad + \sum_k\frac{a_{k}}{a_i}\left(\phi_i-\phi_k\right)\sigma_{i,k}^{(-)}\Bigg]\; .
\end{align}
Again we see the ubiquitous $\sigma_{i,j}^{(\pm)}$ parameters (c.f. Eqs. \eqref{sig_p} and \eqref{sig_m}) rearing their heads in the off-diagonal terms, while the diagonal terms take the same form as they do around proportional solutions (c.f. Eq. \eqref{W_comps_prop}).

To solve the field equations, two things must happen: first, the diagonal ($\delta^\mu_\nu$) part of the $W$-tensors must specify the value of the effective cosmological constant via Eqs. \eqref{propsols}; second, the off-diagonal ($l^\mu l_\nu$) components of the $W$-tensors must vanish. The latter is accomplished by setting either $\phi_i=\phi_j$ or $\sigma_{i,j}^{(+)}=0$ along each interaction link. This can be achieved in three distinct ways:
\begin{enumerate}
    \itemsep0em
    \item Set \emph{all} $\phi_i=\phi_j$. All metrics are proportional, and one is free to use the results of section \ref{sec:prop} to determine the perturbation structure.

    \item Set \emph{all} $\sigma_{i,j}^{(+)}=0$ (remember, the $\sigma's$ are proportional to one another so this also means all $\sigma_{j,i}^{(-)}=0$). None of the metrics are proportional.

    \item Set $\phi_i=\phi_j$ along some interaction links but $\sigma_{i,j}^{(+)}=0$ along others. Some, but not all, of the metrics are proportional.
\end{enumerate}
The latter two branches require one to fine-tune the $\beta_m^{(i,j)}$ interaction coefficients to get a valid solution, and we argued in \cite{BHs_multigrav,BH2} that they should actually be pathological anyway, as setting $\sigma_{i,j}^{(+)}=0$ leads to one of the heavy spin-2 fields becoming massless asymptotically far from the black hole, in violation of Boulanger \emph{et al}'s no-go theorem \cite{no_interacting_massless_gravitons}. Nevertheless, until this pathology is explicitly confirmed, let us determine the structure of the perturbations for non-proportional solutions (option 2). Setting all the $\sigma$'s equal to 0 in Eqs. \eqref{BHperts1} and \eqref{BHperts2} means that the contribution of the mass terms to the linearised field equations \eqref{GeneralLinearisedFieldEqs} take the form:
\begin{align}
    &[\mathcal{B}_{i,j}^{(+)}]^{\mu\;\;\alpha}_{\;\;\nu\;\;\beta} (\bar{S}_{i\rightarrow j})^\beta_{\;\lambda}\tensor{\chi}{^{\lambda}_\alpha} \nonumber
    \\
    &=\mathcal{C}_{i,j}^{(+)}(\phi_i-\phi_j) \left[2\tensor{\chi}{^{(\mu}_\alpha}l^{\alpha)}l_\nu + \delta^\mu_\nu \tensor{\chi}{^{\alpha}_\beta}l^\beta l_\alpha - \tensor{\chi}{^{\alpha}_\nu} l^\mu l_\alpha\right] \; ,\label{dWnonpropBH1}
    \\
    &[\mathcal{B}_{i,k}^{(-)}]^{\mu\;\;\alpha}_{\;\;\nu\;\;\beta}(\bar{S}_{i\rightarrow k})^\beta_{\;\lambda}\tensor{\chi}{^{\lambda}_\alpha} \nonumber
    \\
    &= \mathcal{C}_{i,k}^{(-)} (\phi_i-\phi_k)\left[2\tensor{\chi}{^{(\mu}_\alpha}l^{\alpha)}l_\nu + \delta^\mu_\nu \tensor{\chi}{^{\alpha}_\beta}l^\beta l_\alpha - \tensor{\chi}{^{\alpha}_\nu} l^\mu l_\alpha\right] \; ,\label{dWnonpropBH2}
\end{align}
where we have defined the parameters:
\begin{align}
    \mathcal{C}_{i,j}^{(+)} &= \frac{a_j^2}{a_i^2}\beta_2^{(i,j)} + \frac{a_j^3}{a_i^3}\beta_3^{(i,j)} \; ,\label{Ap}
    \\
    \mathcal{C}_{i,k}^{(-)} &= \frac{a_k^2}{a_i^2}\beta_2^{(k,i)} + \frac{a_k}{a_i}\beta_3^{(k,i)} \; ,\label{Am}
\end{align}
related by:
\begin{equation}\label{BH_Aparams_relation}
    \mathcal{C}_{i,j}^{(+)} = \left(\frac{a_j}{a_i}\right)^4 \mathcal{C}_{j,i}^{(-)} \; .
\end{equation}
Note that $\beta_1$ is not present in these expressions, as we have used $\sigma_{i,j}^{(\pm)}=0$ to express it in terms of $\beta_2$ and $\beta_3$.

Now we can specify our background further to the one used in \cite{GL_unified,Stability_nonbidiag,BH2}, namely, $D=4$ multi-Schwarzschild-(A)dS in (advanced) Eddington-Finkelstein coordinates, where the various functions from Eq. \eqref{KdSMultiGrav} are explicitly:
\begin{align}
    g^{(\Lambda)}_{\mu\nu}\dd x^\mu\dd x^\nu &= - \left(1-\frac{\Lambda}{3} r^2\right)\dd v^2 + 2\dd v \dd r+ r^2\dd\Omega_2^2 \; ,\label{dS_EF_coords}
    \\
    g^{(\Lambda)\mu\nu}\partial_\mu\partial_\nu &= 2\partial_v\partial_r + \left(1-\frac{\Lambda}{3}r^2\right)\partial_r^2 + \frac{\partial_\theta^2+\csc^2\theta\partial_\phi^2}{r^2}\label{dS_EF_coords_inv}
    \\
    l_\mu \dd x^\mu &= \dd v \; , 
    \\
    l^\mu\partial_\mu &= \partial_r \; ,
    \\
    \phi_i &= \frac{r_{s,i}}{2r} \; ,\label{phi_EF_coords}
\end{align}
with $r_{s,i}$ the Schwarzschild radius associated to each metric. With these choices, the linearised Lorentz constraints \eqref{BH_lorentz} become:
\begin{align}
    [\tensor{\omega}{_{mn}}]_{i,j} &= 0 \; ,
    \\
    4[\tensor{\omega}{_{0m}}]_{i,j} &= -\frac{1}{4r}(r_{s,i}-r_{s,j})[\tensor{\delta g}{_{1m}}]_{i,j} \; .
\end{align}
As in the cosmological case, the spatial components are solved trivially but the mixed space-time components are not. 

Regarding the metric equations, substituting into Eqs. \eqref{dWnonpropBH1} and \eqref{dWnonpropBH2}, then into Eq. \eqref{GeneralLinearisedFieldEqs}, one finds that the linearised field equations around this background are:
\begin{align}\label{LinearisedEFEsNonProp}
    &\tensor{\bar{\mathcal{E}}}{^{(i)\mu}_\nu^\alpha_\beta}\tensor{\delta g}{^{(i)\beta}_\alpha} - \Lambda\left(\tensor{\delta g}{^{(i)\mu}_\nu}-\frac12\delta^\mu_\nu\delta g^{(i)}\right) \nonumber
    \\
    &\qquad+ \frac{\kappa_i a_i^2}{2}\left[\sum_j [\chi_{i,j}^{(+)}]^\mu_{\;\nu} + \sum_k [\chi_{i,k}^{(-)}]^\mu_{\;\nu}\right] = 0 \; ,
\end{align}
defining the tensors:
\begin{equation}
    [\chi_{i,j}^{(+)}]^\mu_{\;\nu} = -\left(\frac{a_j}{a_i}\right)^4 [\chi_{j,i}^{(-)}]^\mu_{\;\nu} = \mathcal{C}_{i,j}^{(+)}(\phi_i-\phi_j)[\Delta_{i,j}]^\mu_{\;\nu} \; ,
\end{equation}
where the matrix $\Delta_{i,j}$ has components:
\begin{equation}
    \Delta_{i,j} =
    \begin{bmatrix}
        0 & 0 & 0 & 0
        \\
        [\tensor{\delta g}{^2_2}]_{i,j} + [\tensor{\delta g}{^3_3}]_{i,j} & 0 & -[\tensor{\delta g}{^0_2}]_{i,j}  & -[\tensor{\delta g}{^0_3}]_{i,j} 
        \\
        -[\tensor{\delta g}{^2_1}]_{i,j} & 0 & [\tensor{\delta g}{^0_1}]_{i,j}  & 0
        \\
        -[\tensor{\delta g}{^3_1}]_{i,j} & 0 & 0 & [\tensor{\delta g}{^0_1}]_{i,j}
    \end{bmatrix} \; ,\label{nonprop_int_matrix}
\end{equation}
One should note that \emph{none} of the non-trivial Lorentz parameters enter Eq. \eqref{nonprop_int_matrix} owing to the form of the inverse metric \eqref{dS_EF_coords_inv} -- one has:
\begin{align}
    [\tensor{\delta g}{^0_m}]_{i,j}-4[\tensor{\omega}{^0_m}]_{i,j} &= [\delta g_{1m}]_{i,j} - 4[\omega_{1m}]_{i,j} \nonumber
    \\
    &= [\delta g_{1m}]_{i,j} \nonumber
    \\
    &= [\tensor{\delta g}{^0_m}]_{i,j} \; .
\end{align}
Therefore, on this particular background, simply knowing the form of the background $\bar{S}_{i\rightarrow j}$ matrices is enough to fully specify the form of the linearised field equations.

Since the background is non-rotating, it is spherically symmetric. This means that the metric perturbations can be decomposed into a complete basis of tensor spherical harmonics. In Fourier space, the decomposition reads:
\begin{equation}\label{ax_pol_decomposition}
    \delta\gi{i}(v,r,\theta,\phi) = \sum_{l,m} \frac{a_i^2}{\sqrt{2\pi}} \int_{-\infty}^{\infty}\dd\omega\, e^{-\text{i}\omega v} \delta\tilde{g}^{(i)l m}_{\mu\nu}(\omega,r,\theta,\phi) \; ,
\end{equation}
where:
\begin{equation}
    \delta\tilde{g}^{(i)lm}_{\mu\nu} = \delta\tilde{g}^{(i)\text{ax},lm}_{\mu\nu} + \delta\tilde{g}^{(i)\text{pol},lm}_{\mu\nu} \; .
\end{equation}
The superscript ``ax'' stands for axial (i.e. parity-odd) perturbations, and the superscript ``pol'' stands for polar (i.e. parity-even) perturbations. The spherical symmetry ensures that the field equations for the axial and polar perturbations decouple, as do those for different harmonic indices $l$. The explicit expressions for the axial and polar perturbation matrices are (suppressing $(i)$ indices, and indicating symmetric components with asterisks) \cite{Kerr_instability_bigravity,Stability_nonbidiag}:
\begin{widetext}
\begin{align}
    \delta\tilde{g}^{\text{ax},lm}_{\mu\nu} &=\label{hax}
    \begin{bmatrix}
        0 & 0 & h_0^{lm}(\omega,r)\csc\theta\partial_\phi Y_{lm} & - h_0^{lm}(\omega,r)\sin\theta\partial_\theta Y_{lm}
        \\
        * & 0 & h_1^{lm}(\omega,r)\csc\theta\partial_\phi Y_{lm} & - h_1^{lm}(\omega,r)\sin\theta\partial_\theta Y_{lm}
        \\
        * & * & -h_2^{lm}(\omega,r)\csc\theta X_{lm} & h_2^{lm}(\omega,r)\sin\theta Z_{lm}
        \\
        * & * & * & h_2^{lm}(\omega,r)\sin\theta X_{lm}
    \end{bmatrix} \; ,
    \\
    \delta\tilde{g}^{\text{pol},lm}_{\mu\nu} &= \begin{bmatrix}
        H_0^{lm}(\omega,r)Y_{lm} & H_1^{lm}(\omega,r)Y_{lm} & \eta_0^{lm}(\omega,r)\partial_\theta Y_{lm} & \eta_0^{lm}(\omega,r)\partial_\phi Y_{lm}
        \\
        * & H_2^{lm}(\omega,r)Y_{lm} & \eta_1^{lm}(\omega,r)\partial_\theta Y_{lm}& \eta_1^{lm}(\omega,r)\partial_\phi Y_{lm}
        \\
        * & * & r^2\Big[\begin{smallmatrix}K^{lm}(\omega,r)Y_{lm}
        \\+G^{lm}(\omega,r)Z_{lm}\end{smallmatrix}\Big] & r^2 G^{lm}(\omega,r) X_{lm}
        \\
        * & * & * & r^2\sin^2\theta\Big[\begin{smallmatrix}K^{lm}(\omega,r)Y_{lm} \\ - G^{lm}(\omega,r)Z_{lm}\end{smallmatrix}\Big]
    \end{bmatrix} \; .\label{hpol}
\end{align}
\end{widetext}
Here, $Y_{lm}(\theta,\phi)$ are the ordinary (scalar) spherical harmonics, the functions $X_{lm}(\theta,\phi)$ and $Z_{lm}(\theta,\phi)$ are given by:
\begin{align}
    X_{lm}(\theta,\phi) &= 2\partial_\phi\left(\partial_\theta Y_{lm}-\cot\theta Y_{lm}\right) \; ,\label{Xlm}
    \\
    Z_{lm}(\theta,\phi) &= \partial_\theta^2 Y_{lm}-\cot\theta\partial_\theta Y_{lm} -\csc^2\theta\partial_\phi^2 Y_{lm} \; ,\label{Zlm}
\end{align}
and all of the remaining functions of $(\omega,r)$ are free.

Raising an index with the background metric \eqref{KdSMultiGrav}, written in the Eddington-Finkelstein coordinates of Eqs. \eqref{dS_EF_coords}--\eqref{phi_EF_coords}, then substituting into Eq. \eqref{nonprop_int_matrix}, one finds that the interaction matrix $\Delta_{i,j}$, in Fourier space, explicitly reads as follows:
\begin{widetext}
\begin{equation}\label{PerturbationMatrixNonProp}
    \Delta_{i,j} =
    \begin{bmatrix}
            0 & 0 & 0 & 0
            \\
            2[K^{lm}]_{i,j} Y_{lm} & 0 & -\Big(\begin{smallmatrix}
                \csc\theta\partial_\phi Y_{lm}[h_1^{lm}]_{i,j}
                \\+\partial_\theta Y_{lm}[\eta_1^{lm}]_{i,j}\end{smallmatrix}\Big) & \Big(\begin{smallmatrix}\sin\theta\partial_\theta Y_{lm}[h_1^{lm}]_{i,j} \\- \partial_\phi Y_{lm}[\eta_1^{lm}]_{i,j}\end{smallmatrix}\Big)
            \\
            -\Big(\begin{smallmatrix}
            \csc\theta\partial_\phi Y_{lm}[h_1^{lm}]_{i,j}\\+\partial_\theta Y_{lm}[\eta_1^{lm}]_{i,j}\end{smallmatrix}\Big) & 0 & [H_2^{lm}]_{i,j} Y_{lm} & 0
            \\
            \frac{1}{r^2\sin^2\theta}\Big(\begin{smallmatrix}\partial_\theta Y_{lm}[h_1^{lm}]_{i,j}\\-\csc\theta\partial_\phi Y_{lm}[\eta_1^{lm}]_{i,j}\end{smallmatrix}\Big) & 0 & 0 & [H_2^{lm}]_{i,j} Y_{lm}
        \end{bmatrix} \; .
\end{equation}
\end{widetext}
This is precisely the perturbation matrix around non-proportional multi-Schwarzschild black holes that was derived in \cite{GL_unified,Stability_nonbidiag,BH2}  (up to the canonical normalisation of the metric perturbations implicit in Eq. \eqref{DiffMetPerts}, which we have included here but which was not included in those works). Everything works as intended.

\bibliographystyle{JHEP}
\bibliography{bibliography.bib}
\end{document}